\newif\ifdraft\drafttrue
\newif\iffull\fulltrue

\documentclass[bsl,reqno]{asl}

\def\copyrightyear{2018}

\def\clap#1{\hbox to 0pt{\hss#1\hss}}

\def\mathclap{\mathpalette\mathclapinternal}

\def\mathclapinternal#1#2{%
\clap{$\mathsurround=0pt#1{#2}$}}

\usepackage{amssymb}
\usepackage{url}
\usepackage{yfonts}
\usepackage{enumitem}           
\usepackage{array}

\usepackage{local}

\usepackage{tikz}
\usetikzlibrary{arrows,shapes,positioning,chains}

\newcommand{\qedhere}{%
\global\qedneededfalse%
\hfill \qed
}

\newtheorem{abscounter}{abscounter}

\newtheorem{theorem}[abscounter]{Theorem}
\newtheorem{lemma}[abscounter]{Lemma}
\newtheorem{corollary}[abscounter]{Corollary}

\theoremstyle{definition}
\newtheorem{definition}[abscounter]{Definition}

\newcommand{\rn}[1]{\textsc{#1}}

\newcommand{\true}{\ensuremath{\mathfrak{t}}}
\newcommand{\false}{\ensuremath{\mathfrak{f}}}

\newcommand{\Implies}{\ensuremath{\Rightarrow}}
\newcommand{\END}{\ensuremath{\mathsf{end}}\xspace}
\newcommand{\X}{\ensuremath{\mathop{\circ}}} 
\newcommand{\WX}{\ensuremath{\mathop{\bullet}}}
\newcommand{\A}{\ensuremath{\mathop{\square}}}
\newcommand{\E}{\ensuremath{\mathop{\lozenge}}}
\newcommand{\W}{\ensuremath{\mathrel{\mathcal{W}}}}
\newcommand{\U}{\ensuremath{\mathrel{\mathcal{U}}}}

\newcommand{\K}{\ensuremath{\mathsf{K}}}
\newcommand{\Kprime}{\ensuremath{\mathsf{K}{'}}}

\newcommand{\F}{\ensuremath{\mathcal{F}}\xspace}
\renewcommand{\P}{\ensuremath{\mathcal{P}}\xspace}
\newcommand{\Q}{\ensuremath{\mathcal{Q}}\xspace}
\newcommand{\Z}{\ensuremath{\mathcal{Z}}\xspace}
\newcommand{\G}{\ensuremath{\mathcal{G}}\xspace}
\newcommand{\Phat}{\ensuremath{\widehat{\mathcal{P}}}\xspace}
\newcommand{\Qhat}{\ensuremath{\widehat{\mathcal{Q}}}\xspace}

\newcommand{\comps}{\ensuremath{\mathsf{comps}}\xspace}
\newcommand{\assigns}{\ensuremath{\mathsf{assigns}}\xspace}
\newcommand{\PNP}{\ensuremath{\mathsf{PNP}}\xspace}
\newcommand{\pos}{\ensuremath{\mathsf{pos}}\xspace}
\newcommand{\negs}{\ensuremath{\mathsf{neg}}\xspace}
\newcommand{\Var}{\ensuremath{\mathsf{Var}}\xspace}

\title{Injecting finiteness to prove \\ completeness for finite linear temporal logic}
\gdef\shorttitle{Injecting finiteness for LTLf completeness}

\begin{document}

\author{Eric Campbell}
\revauthor{Campbell, Eric}
\address{Cornell University \\ Ithaca, NY, USA}
\email{ehc86@cornell.edu}

\author{Michael Greenberg}
\revauthor{Greenberg, Michael}
\address{Stevens Institute of Technology \\ Hoboken, NJ, USA}
\email{michael.greenberg@stevens.edu}

\begin{abstract}
Temporal logics over finite traces are not the same as temporal logics
over potentially infinite
traces~\cite{Baier:2006:PFT:1597538.1597664,IntroLDLf_2012,Insensitivity_2014}.
Ro\c{s}u first proved completeness for linear temporal logic on finite
traces (\LTLf) with a novel coinductive axiom~\cite{CoinductiveLTLf_2016}.
We offer a different proof, with fewer, more conventional
axioms. Our proof is a direct adaptation of Kr\"oger and Merz's
Henkin-Hasenjaeger-style proof~\cite{KrogerMerz_LTL_2008}. The essence of our adaption is that
we ``inject'' finiteness: that is, we alter the proof structure to
ensure that models are finite. We aim to present a thorough,
accessible proof.
\end{abstract}

\maketitle

\section{Introduction}
\label{sec:intro}

Temporal logics have proven useful in a remarkable number of applications, in
particular reasoning about reactive systems. To accommodate the nonterminating
nature of such systems, temporal logics have used a possibly infinite model of
time.
For nearly thirty years after Pnueli's seminal work~\cite{Concurrent_1977}, the
prevailing wisdom held that proofs about infinite-trace temporal logics were
sound for finite models of time.
Researchers have recently overturned that conventional wisdom: some formulae are
valid only in finite
models (and vice versa)~\cite{Baier:2006:PFT:1597538.1597664,IntroLDLf_2012,Insensitivity_2014}.

Having realized that finite temporal logics differ from (possibly) infinite ones,
we may wonder: how do these finite temporal logics behave? What are their model
and proof theories like? Can we adapt existing metatheoretical techniques from
infinite settings, or must we come up with new ones?
Reworking the model theory of temporal logics for finite time is an uncomplicated
exercise: the standard model is a (possibly infinite) sequence of valuations on
primitive propositions; to consider only finite models, simply restrict the model
to finite sequences of valuations.
The proof theory is more challenging.
In practice, it is sufficient to (a) add an axiom indicating that the end of
time eventually comes, (b) add an axiom to say what happens when the end of time
arrives, and (c) to relax (or strengthen) axioms from the infinite logic that
may not hold in finite settings.
For an example of (c), consider \LTLf. It normally holds that the next modality
commutes with implication, i.e., $\X (\phi \Implies \psi) \Iff (\X \phi \Implies \X \psi)$, i.e., in the next moment $\phi$ implies $\psi$ iff $\phi$ in the next moment implies $\psi$ in the next moment;
in a finite setting, we must relax the if-and-only-if to merely the
left-to-right direction.

Once we settle on a set of axioms, what does a proof of deductive
completeness look like?
We believe that it is possible to adapt existing techniques for
infinite temporal logics to finite ones \emph{directly}. As evidence,
we offer a proof of completeness for linear temporal logic over finite
traces (\LTLf) with a a conventional structure: we define a graph of
\emph{positive-negative pairs} of formulae (PNPs), following Kr\"oger
and Merz's presentation~\cite{KrogerMerz_LTL_2008}.
The only change we make to their construction is that when we prove our
satisfiability lemma---the core property relating the PNP graph to
provability---we ``inject'' finiteness, adding a formula that guarantees a
finite model.

\smallskip

We claim the following contributions:
\begin{itemize}
\item Evidence for the claim that the metatheory for infinite temporal logics readily
      adapts to finite temporal logics by means of \emph{injecting finiteness}
      (Section~\ref{sec:related} situates our work;
       Section~\ref{sec:model} explains our model of finite time).
\item A proof of deductive completeness for linear temporal logic on finite
      traces (\LTLf; Section~\ref{sec:ltlf}) with fewer axioms than any prior
      proof~\cite{CoinductiveLTLf_2016}.
\end{itemize}

\section{Related work}
\label{sec:related}

Pnueli~\cite{Concurrent_1977} proved his temporal logic programs to be sound and
complete over traces of ``discrete systems'' which may or may not be finite;
Lichtenstein et al.~\cite{PastTimeLTL_1985} extended LTL with past-time
operators and allowed more explicitly for the possibility of finite or
infinite traces.

Baier and McIlraith were the first to observe that some formulae are only valid
in infinite models, and so \LTLf and other `truncated' finite temporal logics
differ from their infinite
originals~\cite{Baier:2006:PFT:1597538.1597664}. Ro\c{s}u~\cite{CoinductiveLTLf_2016}
offers a translation from \LTLf to \LTL that perserves satisfiability of
formulae, but makes no claims about the inverse translation.  De Giacomo and
Vardi showed that satisfiability and validity were PSPACE-complete for these
finite logics, relating \LTLf and linear dynamic logic (\LDLf) to other logics
(potentially infinite LTL, FO$[<]$, star-free regular expressions, MSO on finite
traces)~\cite{IntroLDLf_2012}; later, de Giacomo et al. were able to directly
characterize when \LTLf and \LDLf formulae are sensitive to
infiniteness~\cite{Insensitivity_2014}. De Giacomo and Vardi have also studied
the synthesis problem for our logic of
interest~\cite{de2015synthesis,DeGiacomo:2016:LFL:3060621.3060766}.
Most recently, D'Antoni and Veanes offered a decision procedure for
MSO on finite sequences, but without a deductive completeness
result~\cite{DAntoni:2017:MSL:3009837.3009844}.

Ro\c{s}u~\cite{CoinductiveLTLf_2016} was the first to show a deductive
completeness result for a finite temporal logic: he showed \LTLf is
deductively complete by replacing the induction axiom with a
\emph{coinduction} axiom \rn{coInd}: if $\vdash \WX \phi \Implies
\phi$ then $\vdash \A \phi$.\footnote{In Ro\c{s}u's paper, empty circles
  mean ``weak next'' and filled ones mean ``next'', while we follow
  Kr\"oger and Merz and do the reverse---even when quoting Ro\c{s}u~\cite{KrogerMerz_LTL_2008}.}
He shows that \rn{coInd} is equivalent to the combination of the
conventional induction axiom $\rn{Ind}$ (if $\vdash \phi \Implies \WX
\phi$ then $\vdash \phi \Implies \A \phi$) axiom and a finiteness
axiom $\rn{Fin}$, $\vdash \E \WX \bot$.
Ro\c{s}u proves consistency using ``maximally consistent'' worlds,
i.e., in a greatest fixed-point style.

Our goal is to show that existing, conventional methods for infinite
temporal logics suffice for proving that finite temporal logics are
deductively complete.
For \LTLf, we take the conventional inductive framing,
extending Kr\"oger and Merz's axioms with the axiom $\vdash \E \WX \bot$, i.e., $\E \END$ (we call this axiom \rn{Finite}).
Surprisingly, we are able to prove completeness with only six temporal
axioms---one fewer than Ro\c{s}u's seven, though he conjectures his
set is minimal. It turns out that some of his axioms are
consequences of others---his necessitation axiom $\rn{N}_{\A}$ can be proved from
$\rn{N}_{\WX}$ and $\rn{coInd}$ (we use \rn{Induction} and \rn{WkNextStep}, our equivalent of $\rn{N}_{\WX}$, in Lemma~\ref{lem:boxstep}).
Our results for \LTLf show that a smaller axiom set exists. In fact,
we could go still smaller: using Ro\c{s}u's proofs, we can replace
\rn{Finite} and \rn{Induction} with \rn{coInd}, for only five axioms!
Our proof offers a separate contribution, beyond shrinking
the number of axioms needed and giving a thorough, accessible presentation: we follow Kr\"oger and Merz's least fixed-point
construction quite closely, adapting their proof from LTL to \LTLf by
\emph{injecting finiteness}.
The key idea is that existing techniques for infinite systems readily
adapt to finite ones: we can reuse model theory which uses potentially
infinite models so long as we can force the theory to work exclusively with
finite models.

\subsection{Applications}
\label{sec:applications}

For de Giacaomo and Vardi, \LTLf is useful for AI planning
applications~\cite{IntroLDLf_2012,Insensitivity_2014,de2015synthesis,DeGiacomo:2016:LFL:3060621.3060766}.
The second author first encountered finite temporal logics when
designing Temporal NetKAT~\cite{TemporalNetKat_2016}. NetKAT is a
specification language for network configurations~\cite{NetKat_2014}
based on Kleene algebra with tests~\cite{Kozen97kat}. Temporal NetKAT
extends NetKAT with the ability to write and analyze policies using
past-time finite linear temporal logic, e.g., a packet may not arrive
at the server unless it has previously been at the firewall.
Our interest in the deductive completeness of \LTLf comes directly
from the Temporal NetKAT work: the completeness result for Temporal
NetKAT's equivalence relation relies on deductive completeness for
\LTLf.

\section{Modeling finite time}
\label{sec:model}

Our logic, \LTLf, uses a finite model of time: \emph{traces}.
A trace over a fixed set of propositional variables is a (possibly infinite) 
sequence $(\eta_1,\eta_2,\dots)$ where $\eta_i$ is a \emph{valuation}, i.e., a
function establishing the truth value ($\true$ or $\false$) for each propositional variable. We refer
to each valuation as a `state', with the intuition that each valuation represents
a discrete moment in time.
\begin{definition}[Valuations and Kripke structures]
  \label{def:kripke}
  Given a set of variables $\Var$, a \emph{valuation} 
  is a function $\eta : \Var \rightarrow \{ \true, \false \}$. A 
  \emph{Kripke structure} or a \emph{trace} is a finite, non-empty sequence of 
  valuations; we write $\K^n \in \mathsf{Model}_n$ to refer to a model with $n \in \mathbb{N}^{+}$
  valuations, i.e., $\K^n = (\eta_1,\dots,\eta_n)$.
\end{definition}
We particularly emphasize the finiteness of our Kripke structures,
writing $\K^n$ and explicitly stating the number of valuations as a
superscript each time. The number $n$ is not directly accessible in
our logic, though the size of models is observable (e.g., the \LTLf
formula $\X \X \X \top$ is satisfiable only in models with four
or more steps).
Our traces are not only finite, but they are necessarily
non-empty---all \LTLf formulae would trivially hold in empty models.

Suppose we have $\Var = \{ x, y, z \}$. As a first example,
the smallest possible model is one with only one time step, $\K^1
= \eta_1$, where $\eta_1$ is a function from $\Var$ to the
booleans, i.e., a subset of $\Var$. As a more complex example,
consider the following model $\K^4$ with four time steps:
\begin{center}
\begin{tikzpicture}
  \draw (-0.8,0) node {$\K^4 = {}$};
  \draw (0,0) -- (6,0);
  \foreach \i in {0,2,4,6}
    \draw[shift={(\i,0)},color=black] (0pt,3pt) -- (0pt,-3pt);
  \foreach \i in {1,2,3,4}
    \draw[shift={({2*(\i-1)},0)},color=black] (0pt,0pt) -- (0pt,-3pt) node[below]
      {$\eta_\i$};
  \draw[shift={(0,0)},color=black] (0pt,0pt) -- (0pt,3pt) node[above] {$\{x\}$};
  \draw[shift={(2,0)},color=black] (0pt,0pt) -- (0pt,3pt) node[above] {$\{x,y\}$};
  \draw[shift={(4,0)},color=black] (0pt,0pt) -- (0pt,3pt) node[above] {$\emptyset$};
  \draw[shift={(6,0)},color=black] (0pt,0pt) -- (0pt,3pt) node[above] {$\{x,y,z\}$};
\end{tikzpicture}
\end{center}
{\iffull
In the first state, $x$ holds but $y$ and $z$ do not (i.e. $\eta_1(x)
= \true$, but $\eta_1(y) = \eta_1(z) = \false$); then $x$ and $y$
hold; then no propositions hold; and then all primitive propositions
hold.
\fi}

Our logic uses Kripke structures to interpret formulae,
defining a function $\K^n_i
: \mathsf{Formula} \rightarrow \{ \true, \false \}$.
(Put another way: we define a function $\mathsf{interp}
: \mathsf{Model}_n \times \{ 1, \dots,
n \} \times \mathsf{Formula} \rightarrow \{ \true, \false \}$, writing $K^n_i(\phi)$ for $\mathsf{interp}(K^n,i,\phi)$.)
We lift this interpretation function to define validity and
satisfiability.

\begin{definition}[Semantic satisfiability and validity]
  \label{def:satvalid}
  For an interpretation function $\K^n_i
  : \mathsf{Formula} \rightarrow \{ \true, \false \}$, we say for
  $\phi \in \mathsf{Formula}$:
  \begin{itemize}

  \item \emph{$\K^n$ models $\phi$} iff $\K^n_1(\phi) = \true$;

  \item $\phi$ is \emph{satisfiable} iff $\exists \K^n$ such that
    $\K^n$ models $\phi$;

  \item $\K^n \models \phi$ (pronounced ``$\K^n$ satisfies $\phi$'') iff
  $\forall 1 \le i \le n, ~ \K^n_i(\phi) = \true$;
  
  \item $\models \phi$ (pronounced ``$\phi$ is valid'') iff
  $\forall \K^n, \K^n \models \phi$; and

  \item $\F \models \phi$ for $\F \subseteq \mathsf{Formula}$
  (pronounced ``$\phi$ is valid under $\F$'') iff $\forall \K^n,$ if
  $\forall \psi \in \F, ~ \K^n \models \psi$ then $\K^n \models \phi$.

  \end{itemize}
\end{definition}

\section{\LTLf: linear temporal logic on finite traces}
\label{sec:ltlf}

Linear temporal logic is a classical logic for reasoning on potentially infinite traces.
The syntax of linear temporal logic on finite traces (\LTLf) is identical to 
that of its (potentially) infinite counterpart. We define \LTLf as a propositional
logic with two temporal operators (Figure~\ref{fig:ltlfsemantics}).
The propositional fragment comprises:
variables $v$ from some fixed set of propositional variables $\Var$;
the false proposition, $\bot$;
and implication, $\phi \Implies \psi$.
The temporal fragment comprises two operators:
the \emph{next modality}, written $\X \phi$, which means that $\phi$ holds in the next moment of time;
and, \emph{weak until}, written $\phi \W \psi$, which means that $\phi$ holds until either (a) the end of time, or (b) $\psi$ holds.

\begin{figure}[t]
  \hdr{Syntax}{}
  \vspace*{-1em}

  \[ \begin{array}{l@{~~}r@{~}c@{~}lcl}
    \text{Variables} & v    &\in& \Var && \\
    \text{\LTLf formulae}  & \phi, \psi &\in& \LTLf        &::=&
       v \BNFALT \bot \BNFALT \phi \Implies \psi \BNFALT \X \phi \BNFALT \phi \W \psi \\
  \end{array} \]

  \hdr{Encodings}{}
  \vspace*{-1em}

  \[ \begin{array}{r@{~=~}l@{\qquad}r@{~=~}l@{\qquad}r@{~=~}l}
    \neg \phi        & \phi \Implies \bot              &
    \top             & \neg \bot                       &
    \phi \vee \psi   & \neg \phi \Implies \psi         \\[.25em]
    \phi \wedge \psi & \neg (\neg \phi \vee \neg \psi) &
    \END             & \neg \X \top &
    \WX \phi         & \neg \X \neg \phi \\[.25em]
    \A \phi          & \phi \W \bot &
    \E \phi          & \neg \A \neg \phi &
    \phi \U \psi     & \phi \W \psi \wedge \E \psi \\
  \end{array} \]

  \hdr{Semantics}{\fbox{$\K^n_i : \LTLf \rightarrow \{ \true, \false \}$}}
  \begin{align}
    \K^n_i(v) &= \eta_i(v) \label{sem:var}\\
    \K^n_i(\bot) &= \false \label{sem:bot}\\
    \K^n_i(\phi \Implies \psi) &= \begin{cases}
      \true & \K^n_i(\phi) = \false \text{ or } \K^n_i(\psi) = \true \\
      \false & \text{otherwise}
    \end{cases} \label{sem:imp}\\
    \K^n_i(\X \phi) &= \begin{cases}
      \K^n_{i+1}(\phi) & i < n \\
      \false & i = n
    \end{cases} \label{sem:next}\\
    \K^n_i(\phi \W \psi) &= \begin{cases}
      \true & \forall i \le j \le n, ~ \K^n_j(\phi) = \true
              \text{ or} \\
              & \exists i \le k \le n, ~ \K^n_k(\psi) = \true \text{ and } \\
              & \forall i \le j < k, ~ \K^n_j(\phi) = \true \\
      \false & \text{otherwise}
    \end{cases} \label{sem:wuntil}
  \end{align}

  \caption{\LTLf syntax and semantics}
  \label{fig:ltlfsemantics}
\end{figure}

These core logical operators encode a more conventional looking logic
(Figure~\ref{fig:ltlfsemantics}), with the usual logical operators and an
enriched set of temporal operators.
Of these standard encodings, we remark on two in particular: 
$\END$, the end of time, and $\WX \phi$, the \emph{weak next} modality.
In the usual (potentially infinite) semantics, it is generally the case that
$\X \top$ holds, i.e., that the true proposition holds in the next state, i.e.,
that there \emph{is} a next state. But at the end of time, there is no next
state---and so $\X \top$ ought not adhere. In every state \emph{but} the last,
we have $\X \top$ as usual.
We can therefore define $\END = \neg \X \top$---if $\END$ holds, then we must be
at the end of time.
It's worth noting here that negation does not generally commute with the next
modality;\footnote{This is not true in the possibly-infinite semantics, where
$\models \neg \X \phi \Iff \X \neg \phi$} observe that $\neg \X \top$ holds
only at the end of time, but $\X \neg \top$ holds nowhere.
In fact, $\neg \X \neg \phi$ holds at the end of time for every possible $\phi$,
and everywhere else, $\neg$ and $\X$  commute,
i.e. $\neg \X \neg \phi$ holds if and only if $\X \phi$ holds.
Bearing these facts in mind, we define the \emph{weak next} modality as
$\WX \phi = \neg \X \neg \phi$. Weak next ($\WX \phi$) is \emph{insensitive} to
the end of time and strong next ($\X \phi$) is \emph{senstitive} to the end of time.
To realize these intuitions, we must define our model.

\begin{figure}[t]
  {\flushleft\textbf{Axioms}
   \[\begin{array}{@{}l@{\qquad\qquad\qquad}r@{}}
  {\text{all propositional tautologies}} & \rn{Taut} \\[.5em]

  {\vdash \WX(\phi \Implies \psi) \Iff (\WX \phi \Implies \WX \psi)}
  & \rn{WkNextDistr} \\[.5em]

  {\vdash \END \Implies \neg \X \phi} & \rn{EndNextContra} \\[.5em]

  {\vdash \E \END} & \rn{Finite} \\[.5em]

  {\vdash \phi \W \psi \Iff \psi \vee (\phi \wedge \WX(\phi \W \psi))}
  & \quad \rn{WkUntilUnroll} \\[.5em]

  \displaystyle \frac{\vdash \phi}{\vdash \WX \phi} & \rn{WkNextStep} \\[1.5em]

  \displaystyle \frac{\vdash \phi \Implies \psi \quad \vdash \phi \Implies \WX \phi}
       {\vdash \phi \Implies \A \psi} & \rn{Induction} \\
  \end{array} \]

  \[ \F \vdash \phi \text{ iff assuming $\vdash \psi$ for each $\psi \in \F$ we have } \vdash \phi \]}
  {\flushleft\textbf{Consequences}
   \[\begin{array}{@{}l@{~~}l@{\qquad}l@{~~}l@{}}
  \vdash \neg(\X \top \wedge \X \bot) & Lemma~\ref{lem:nextcontra} &
  \vdash \neg \X \phi \Iff \END \vee \X \neg \phi & Lemma~\ref{lem:commnegnext} \\[.4em]
  \vdash \WX \phi \Iff \X \phi \vee \END & Lemma~\ref{lem:nextwknext} &
  \vdash \neg \END \wedge \WX \neg \phi \Implies \neg \WX \phi & Lemma~\ref{lem:wknextnegend} \\[.4em]
  \vdash \WX(\phi \wedge \psi) \Iff \WX \phi \wedge \WX \psi & Lemma~\ref{lem:wknextdistconj} &
  \vdash \neg \WX \phi \Implies \WX \neg \phi & Lemma~\ref{lem:wknextneg} \\
  \vdash \A \phi \Iff \phi \wedge \WX \A \phi & Lemma~\ref{lem:alwaysunroll} &
  \displaystyle \frac{\vdash \phi}{\vdash \A \phi} & Lemma~\ref{lem:boxstep} \\
  \end{array} \]}
  
  \caption{\LTLf proof theory}
  \label{fig:ltlfaxioms}
\end{figure}

The simple, standard model for LTL is a possibly-infinite \emph{trace}; we
restrict ourselves to finite traces (Definition~\ref{def:kripke}).
Given a Kripke structure $\K^n$, we assign a truth value to a proposition $\phi$
at time step $1 \le i \le n$ with the function $\K^n_i(\phi)$, defined as a 
fixpoint on formulae.
The definitions for $\K^n_i$ in the propositional fragment are straightforward 
implementations of the conventional operations.
The definitions for $\K^n_i$ in the temporal fragment also assign the usual
meanings, being mindful of the end of time.
When there is no next state, the formula $\X \phi$ is necessarily false; 
when there is no next state, the formula $\phi \W \psi$ degenerates into 
$\phi \vee \psi$.
Why? Suppose we are at the end of time; 
one of two cases adheres. Either we have $\phi$ until the end of time (which is now!),
or we have $\psi$ and have satisfied the until.
We can verify our earlier intuitions about $\END$ and
$\WX \phi$. Observe that $\K^n_i(\END) = \true$ exactly when $i = n$;
similarly, $\K^n_i(\WX \top) = \true$ for all $1 \le i \le n$.

By way of example, consider $\K^4$ from Section~\ref{sec:model}. We
have $\K^4 \models y \Implies x$, because $\K^4_i(y \Implies x) = \true$
for all $i$, i.e., whenever $y$ holds, so does $x$. Similarly, $\K^4$ models $x \W
y$ with the $k$ in the existential equal to 2; we have $\K^4$ models $z \W x$, too,
but trivially with $k=1$. The formula $\A z$ doesn't hold in any state of
$\K^4$, but $\K^4 \models \E z$.
%
We prove a semantic deduction theorem appropriate to our setting:
rather than producing a bare implication, deduction produces an
implication whose premise is under an `always' modality.

\begin{theorem}[Semantic deduction]
  \label{thm:semanticdeduction}
  $\F \cup \{ \phi \} \models \psi$ iff $\F \vdash \A \phi \Implies
  \psi$.
  \begin{proof}
    We prove each direction separately.
    From left to right, suppose $\F \cup \{\phi\} \models
      \psi$. Let $\K^n$ be given such that $\K^n \models \chi$ for all
      $\chi \in \F$. We show that $\K^n_i(\A \phi \Implies \psi)$
      for all $i$.

      Let an $i$ be given. If $\K^n_i(\A \phi) = \false$, we are done
      immediately---so suppose $\K^n_j(\A \phi) = \true$ for all $j \ge
      i$. It remains to be seen that $\K^n(\psi) = \true$ for all $j
      \ge i$.
      Let $j$ be given. We can extract a smaller Kripke structure from $K^n$; call it $\Kprime^{n-i} =
      (\eta_i,\dots,\eta_{n-i})$, noting that $\Kprime^{n-i}_k
      = \K^{n}_{k-1+i}$ for all $1 \leq k \leq m-i$. Then, our assumption that
      $\K^n \models \chi$ for all $\chi \in \F \cup \{\phi\}$ implies
      $\Kprime^{n-i} \models \chi$ for all $\chi \in \F \cup {\phi}$. We already
      assumed that $\F \vdash \{\phi\} \models \psi$, so we can conclude that
      $\Kprime^{n-i} \models \psi$.
      Hence, $\Kprime^{n-i}$ assigns $\true$ to $\A \phi \Implies \psi$, and so
      $\K^n_i(\A \phi \Implies \psi) = \true$.

   From right to left, suppose $\F \models \A \phi \Implies \psi$. Let $\K^n$
      be given such that $\K^n \models \chi$ for all
      $\chi \in \F \cup \{ \phi \}$. We must show that $\K^n_i(\psi)$ for all
      $i$. Since $\K^n \models \chi$ for all $\chi \in \F$, then
      $\K^n_i(\A \phi \Implies \psi) = \true$ by assumption. Furthermore, we
      know that $\K^n \models \phi$, i.e., $\K^n_j(\phi) = \true$ for all $i \le
      j \le n$. Then
      $\K^n_i(\A \phi) = \true$ by definition, so the implication in the assumption cannot
      hold vacuously: so $\K^n_i(\psi) = \true$ as
      desired.
  \end{proof}
\end{theorem}

For our axioms (Figure~\ref{fig:ltlfaxioms}), we adapt Kr\"oger and Merz's
presentation~\cite{KrogerMerz_LTL_2008}.
Two axioms are new: 
\rn{Finite} says that time will eventually end; 
\rn{EndNextContra} says that at the end of time, there is no next state.
Other axioms are lightly adapted: wherever one would ordinarily use the (strong) next 
modality, $\X \phi$, we instead use weak next, $\WX \phi$.
Using strong next would be unsound in finite models. We can,
however, characterize the relationship between the next modality,
negation, and the end of time (``Consequences'' in
Figure~\ref{fig:ltlfaxioms} and Section~\ref{sec:consequences}).

Ro\c{s}u proves completeness with a slightly different set of axioms, replacing
\rn{Finite} and \rn{Induction} with a single \emph{coinduction} axiom he calls
\rn{coInd}: \[ \frac{\vdash \WX \phi \Implies \phi}{\vdash \A \phi} \]
He proves that \rn{coInd} is equivalent to the conjunction of \rn{Finite} and
\rn{Induction}, so it does not particularly matter which axioms we choose. 
In order to emphasize how little must change to make our logic finite,
we keep our presentation as close to Kr\"oger and Merz's as
possible.\footnote{They use a slightly less-expressive logic, omitting $\W$ and
$\U$. We extend their methodology to include these operators. }

\subsection{Soundness}

Proving that our axioms are sound is, as usual, relatively
straightforward: we verify each axiom in turn.
\begin{theorem}[\LTLf soundness]
  \label{thm:ltlfsoundness}
  If $\vdash \phi$ then $\models \phi$.
  \begin{proof}

  By induction on the derivation of $\vdash \phi$. Our proof refers to
  the various cases in the definition of the model
  (Figure~\ref{fig:ltlfsemantics}).
    \begin{enumerate}[align=left]
    \item[(\rn{Taut})] As for propositional logic.
    \item[(\rn{WkNextDistr})] We have $\vdash \WX(\phi \Implies \psi)
      \Iff (\WX \phi \Implies \WX \psi)$. To show validity in the model,
      let $\K^n$ be given. We show that $\K^n$ assigns true to the
      left-hand side iff it assigns true to the right-hand side.
      \[ \begin{array}{rcl}
               \multicolumn{3}{l}{\K^n \models \WX (\phi \Implies \psi)} \\
        &\text{iff}& \forall 1 \le i \le n,~ \K^n_i(\WX (\phi \Implies \psi)) = \true \\
        &\text{iff}& \forall 1 \le i \le n-1,~ \K^n_{i+1}(\phi \Implies \psi) = \true \\
        &\text{iff}& \forall 1 \le i \le n-1,~ \K^n_{i+1}(\phi) = \false \text{ or } \K^n_{i+1}(\psi) = \true \\
        &\text{iff}& \forall 1 \le i \le n,~ \K^n_{i}(\WX \phi) = \false \text{
               or } \K^n_{i}(\WX \psi) = \true \\
              \multicolumn{3}{r}{\text{where } \K^n_n(\WX \psi) = \true \text{ trivially}} \\
        &\text{iff}& \forall 1 \le i \le n,~ \K^n_{i}(\WX \phi \Implies \WX \psi) \\
        &\text{iff}& \K^n \models \WX \phi \Implies \WX \psi
      \end{array} \]

      By unfolding the encodings of logical operators, we can derive that
      $\K^n \models \WX(\phi \Implies \psi) \Iff (\WX \phi \Implies \WX \psi)$.

    \item[(\rn{EndNextContra})] We have $\vdash \END \Implies \neg \X
      \phi$; let $\K^n$ be given to show $\K^n \models \END \Implies
      \neg \X \phi$, i.e., that $\K^n$ assigns $\true$ to the given
      formula at each $1 \le i \le n$. Let $i$ be given.

      We have $\K^n_i(\END \Implies \neg \X \phi)$. There are two cases: $i < n$ and $i = n$.
      When $i<n$, we have $\K^n_i(\END) = \K^n_i(\neg \X \top)$. Since $i <
          n$, then $\K^n_i(\X \top) = \K^n_{i+1}(\top)
          = \K^n_{i+1}(\neg \bot) = \true$, and so $\K^n_i(\END) = \false$. The
          implication is thus vacuous: $\K^n_i(\END \Implies \neg \X \phi)
          = \true$, by the first clause of case (\ref{sem:imp}).

      When $i=n$, we have $\K^n_n(\X \phi) = \false$, so
          $\K^n_n(\neg \X \phi) = \true$. Therefore
          $\K^n_i(\END \Implies \neg \X \phi) = \true$, by the second clause of
          case (\ref{sem:imp}).

    \item[(\rn{Finite})] We have $\vdash \E \END$; let $\K^n$ be given
      to show $\K^n \models \E \END$, i.e., that for all $1 \le i \le
      n$, we have $\K^n_i(\E \END) = \true$. Let $i$ be given.

      Unfolding our encodings, we must show that: \[ \K^n_i(\E \END)
      = \K^n_i(\neg \A \neg \END) = \K^n_i(\neg (\neg \END \W \bot)) = \true \]
      By cases (\ref{sem:imp}) and (\ref{sem:bot}), it suffices to show that
      $\K^n_i(\neg \END \W \bot) = \false$. By case (\ref{sem:wuntil}), there
      are two ways for the weak-until to be assigned true; we show that
      neither adheres. First, observe that $\K^n_k(\bot) = \false$ for all
      $1 \le k \le n$, so there is no $k$ to satisfy the second clause of case
      (\ref{sem:wuntil}). Next, observe that when $j = n$, where
      $\K^n_j(\neg \END) = \false$, so the first clause of case
      (\ref{sem:wuntil}) cannot be satisfied. Since neither case holds, we find
      $\K^n_i(\neg \END \W \bot) = \false$.

      \item[(\rn{WkUntilUnroll})] We have $\vdash \phi \W \psi \Iff \psi \vee
      (\phi \wedge \WX(\phi \W \psi))$; let $\K^n$ be given to show that
      $\K^n \models \phi \W \psi \Iff \psi \vee
      (\phi \wedge \WX(\phi \W \psi))$, i.e., that for all $1 \le i \le n$, the
      left-hand side of our formula is assigned true by $\K^n_i$ iff the
      right-hand side is. Let an $i$ be given; we prove each side independently.

      From left to right, we have $\K^n_i(\phi \W \psi) = \true$ iff
        $\forall i \le j \le n, ~ \K^n_j(\phi) = \true$ or $\exists i
        \le k \le n, ~ \K^n_k(\psi) = \true \text{ and } \forall i \le
        j < k, ~ \K^n_j(\phi) = \true$. We go by cases.
        If $\phi$ always holds, then $\K^n_i(\phi) =
          \true$ and $\K^n_{i+1}(\phi \W \psi) = \true$, so $\K^n_i(\psi
          \vee (\phi \wedge \WX(\phi \W \psi))) = \true$.
        If, on the other hand,$\phi$ holds until $\psi$ eventually holds, we ask: is $k=i$?
        If so, then $\K^n_i(\psi) = \true$ implies $\K^n_i(\psi \vee (\phi
          \wedge \WX(\phi \W \psi))) = \true$.
          If not, then $\K^n_{i+1}(\phi \W \psi)$ holds by the second clause of
          (\ref{sem:wuntil}), using our same $k$. Therefore,
          $\K^n_i(\WX(\phi \W \psi)) = \true$ along with $\K^n_i(\phi) = \true$
          (since $j$ can be $i$), so $\K^n_i(\psi \vee
          (\phi \wedge \WX(\phi \W \psi))) = \true$.
        
      From right to left, we have $\K^n_i(\psi \vee (\phi \wedge
        \WX(\phi \W \psi))) = \true$; we must show $\K^n_i(\phi \W
        \psi) = \true$. We go by cases on which side of the
        disjunction holds.
        If $\K^n_i(\psi) = \true$), then $k = i$ witnesses the
            second clause of case (\ref{sem:wuntil}) with $k = i$.
        If $\K^n_i(\phi) = \K^n_i(\WX(\phi \W \psi))=\true$, then we ask: is $i = n$? If so, we
          are done by the first clause of case (\ref{sem:wuntil}).
          If $i < n$, then $\K^n_{i+1}(\phi \W \psi) = \true$. If it holds
          because $\forall i + 1 \le j \le n, ~ \K^n_j(\phi) = \true$, then
          $\K^n_i(\phi) = \true$ completes first clause of case
          (\ref{sem:wuntil}). Otherwise, $\K^n_{i+1}(\phi \W \psi) = \true$,
          because there is some $i + 1 \le k \le n$ such that $\K^n_k(\psi)
          = \true$ and $\forall i+1 \le j < k, ~ \K^n_j(\phi) = \true$. Since we
          also have $\K^n_i(\phi) = \true$, $k$ witnesses the second clause
          of case (\ref{sem:wuntil}).

    \item[(\rn{WkNextStep})] We have $\vdash \WX \phi$; as our IH on
      $\vdash \phi$, we have $\models \phi$, i.e., $\forall \K^n
      \forall 1 \le i \le n, \K^n_i(\phi) = \true$. Let a $\K^n$ be
      given to show $\K^n \models \WX \phi$, i.e., $\forall 1 \le i \le
      n, ~ \K^n_i(\WX \phi) = \true$. Let $i$ be given.
      If $i=n$, then $\K^n_n(\X \neg \phi) = \false$, so
        $\K^n_n(\WX \phi) = \K^n_n(\neg \X \neg \phi) = \true$.
      If $i<n$, then by definition (\ref{sem:next}) $\K^n_i(\X \neg \phi)
        = \K^n_{i+1}(\neg \phi)$.  By the IH, we know that $\K^n_{i+1}(\phi)
        = \true$, so $\K^n_{i+1}(\neg \phi) = \false$, and $\K^n_i(\X \neg \phi)
        = \false$. Therefore $\K^n_i(\WX \phi) = \K^n_i(\neg \X \neg \phi)
        = \true$.
      
    \item[(\rn{Induction})] We have $\vdash \phi \Implies \A \psi$; as our IHs,
      we have (1) $\models \phi \Implies \psi$ and (2)
      $\models \phi \Implies \WX \phi$, i.e., every Kripke structure $\K^n$
      assigns $\true$ to those formulae at every index. Let $\K^n$ be given to
      show that $\forall 1 \le i \le n, ~ \K^n_i(\phi \Implies \A \psi)
      = \true$.  If $\K^n_i(\phi) = \false$, the implication vacuously holds;
      instead consider the case where $\K^n_i(\phi) = \true$.
      Let $k = n-i$; we go by induction on $k$ to show $\K^n_i(\A \psi) = \true$.
      When $k=0$, it must be the case that $n=i$. We have $\K^n_n(\phi) = \true$; by outer IH (1), it
        must be that $\K^n_n(\psi) = \true$, which means that $\K^n_n(\A
        \psi) = \K^n_n(\psi \W \bot) = \true$ by the first clause of
        case (\ref{sem:wuntil}).
 
      When $k=k'+1$, we have here $i < n$. We must show that $\K^n_i(\A
        \psi) = \true$.
        Since $\K^n_i(\phi) = \true$, the outer IHs immediately give
        $\K^n_i(\psi) =_{\text{IH }(1)} \true$ (outer IH on $\vdash \phi \Implies \psi$) and $\K^n_i(\WX \phi) = \true$ (outer IH on $\vdash \phi \Implies \WX \phi$).
        Futhermore, we have $i < n$; unfolding $\WX \phi$
        gives $\K^n_{i+1}(\neg \phi) = \false$, or equivalently
        $\K^n_{i+1}(\phi) = \true$. So $\K^n_{i+1}(\A \psi)
        = \K^n_{i+1}(\psi \W \bot) = \true$ by the inner IH.
        Since $\K^n_j\models \neg \bot$, the above conclusion that
        $\K^n_{i+1}(\psi \W \bot) = \true$ must hold
        because $\forall i+1 \le j \le n, ~ \K^n_j(\psi) = \true$ (as per case
        (\ref{sem:wuntil})). Since $\K^n_i(\psi) = \true$ as well, we have $\forall i \le j \le n ~ \K^n_j(\psi) = \true$; therefore
        $\K^n_i(\A \psi) = \K^n_i(\psi \W \bot) = \true$. \qedhere
  \end{enumerate}
\end{proof}
\end{theorem}

\medskip

We can also prove a deduction theorem for our proof theory analogous
to Theorem~\ref{thm:semanticdeduction}.

\begin{theorem}[Deduction]
  \label{thm:deduction}
  $\F \cup \{ \phi \} \vdash \psi$ iff $\F \vdash \A
  \phi \Implies \psi$.
  \begin{proof}
    From left to right, by induction on the derivation:
    \begin{enumerate}[align=left]
    \item[($\psi$ an axiom or $\psi \in \F$)] We have $\F \vdash \A \phi
      \Implies \psi$ by \rn{Taut}.
      
    \item[(\rn{WkNextStep})] We have $\psi = \WX \chi$. \rn{WkNextStep}
      concludes $\WX \chi$ from $\F \cup \{\phi\} \vdash \chi$, which gives the
      IH of $\F \vdash \A \phi \Implies \chi$. Applying \rn{WkNextStep} to the
      IH, we have $\F \vdash \WX (\A \phi \Implies \chi)$; by \rn{WkNextDistr},
      we have $\F \vdash \WX \A \phi \Implies \WX \chi$. By
      Lemma~\ref{lem:alwaysunroll}, we know that
      $\vdash \A \phi \Implies \phi \wedge \WX \A \phi$, so by \rn{Taut} we have
      $\F \vdash \A \phi \Implies \psi$.
      
    \item[(\rn{Induction})] We have $\psi = \chi \Implies \A \rho$,
      with $\F \cup \{\phi\} \vdash \chi \Implies \rho$ and $F \cup
      \{\phi\} \vdash \chi \Implies \WX \chi$. By the IH, we know that
      $\F \vdash \A \phi \Implies \chi \Implies \rho$ and $\F \vdash
      \A \phi \Implies \chi \Implies \WX \chi$. By \rn{Taut}, we have
      $\F \vdash \A \phi \wedge \chi \Implies \rho$ and $\F \vdash \A
      \phi \wedge \chi \Implies \WX \chi$, so by \rn{Induction} we
      have $\F \vdash \A \phi \wedge \chi \Implies \A \rho$. By
      \rn{Taut}, we find $\F \vdash \A \phi \Implies \chi \Implies \A
      \rho$.
    \end{enumerate}
    From right to left, we have $\F \vdash \A \phi \Implies \psi$ and
    must show $\F \cup \{ \phi \} \vdash \psi$. By \rn{Taut}, we have
    $\F \cup \{\phi\} \vdash \A \phi \Implies \psi$. We must prove that $\F
    \cup \{\phi\} \vdash \A \phi$, which gives $\F \cup
    \{\phi\} \vdash \psi$ via \rn{Taut}.

    By \rn{WkNextStep} and \rn{Taut}, we know that
    $\F \cup \{ \phi \} \vdash \phi \Implies \WX \phi$. We therefore have by
    induction that
    $\F \cup \{ \phi \} \vdash \phi \Implies \A \phi$. By \rn{Taut}, we can
    conclude $\F \cup \{\phi\} \vdash \phi$ and subsequently
    $\F \cup \{ \phi \} \vdash \A \phi$.  \end{proof}
\end{theorem}

\subsection{Consequences}
\label{sec:consequences}

Before proceeding to the proof of completeness, we prove a variety of
properties in \LTLf necessary for the proof:
characterizations of the modality
(Lemmas~\ref{lem:nextcontra},~\ref{lem:nextwknext},
and~\ref{lem:alwaysunroll}) and distributivity over connectives
(Lemmas~\ref{lem:commnegnext},~\ref{lem:wknextnegend},~\ref{lem:wknextdistconj},
and~\ref{lem:wknextneg}). We also derive Ro\c{s}u's necessitation
axiom, $\rn{N}_{\A}$ (Lemma~\ref{lem:boxstep}).

Together Lemma~\ref{lem:wknextnegend} and Lemma~\ref{lem:wknextneg}
completely characterize the relationship between the weak next
modality and negation: we can pull a negation out of a weak
next modality when not at the end; we can push a negation in whether
or not the end has arrived. 

\begin{lemma}[Modal consistency]
  \label{lem:nextcontra}
  $\vdash \neg(\X \top \wedge \X \bot)$
  \begin{proof}
    Suppose for a contradiction that $\vdash \X \top \wedge \X \bot$. We have
    $\vdash \top$ by \rn{Taut}, so $\vdash \WX \top$ by
    \rn{WkNextStep}. But $\WX \top$ desugars to $\neg \X \neg \top$,
    i.e., $\neg \X \bot$---a contradiction.
  \end{proof}
\end{lemma}

\begin{lemma}[Negation of next]
  \label{lem:commnegnext}
  $\vdash \neg \X \phi \Iff \END \vee \X \neg \phi$
  \begin{proof}
    \iffull\else We prove each direction separately using \rn{WkNextDistr}, 
    \rn{EndNextContra}, and \rn{Taut}.\fi
    From left to right, we have $\vdash \END \vee \neg \END$ by the
      law of the excluded middle (\rn{Taut}).  If $\END$ holds, we are
      done. Otherwise, suppose $\vdash \neg \X \phi \wedge \neg \END$; we must
      show $\vdash \X \neg \phi$. By resugaring and \rn{Taut}, we have
      $\vdash \WX \neg \phi$; by \rn{WkNextDistr}, we have
      $\vdash \neg \WX \phi$; by desugaring, we have $\vdash \X \neg \phi$.

    From right to left, the law of the excluded middle yields
      $\vdash \END \vee \neg \END$ (\rn{Taut}). If $\END$ holds, then
      we have $\vdash \neg \X \phi$ by \rn{EndNextContra} immediately.
      So we have $\vdash \neg \END \wedge \X \neg \phi$ and we must show $\vdash \neg
      \X \phi$. By resugaring and \rn{Taut}, we have $\vdash \neg \WX \phi$; by
      \rn{WkNextDistr}, we have $\vdash \WX \neg \phi$. By desugaring and \rn{Taut},
      we have $\vdash \neg \X \phi$ as desired.
  \end{proof}
\end{lemma}

\begin{lemma}[Weak next/next equivalence]
  \label{lem:nextwknext}
  $\vdash \WX \phi \Iff \X \phi \vee \END$
  \begin{proof}
    \iffull\else We prove each direction separately.\fi
    From left to right, suppose $\vdash \WX \phi$. $\WX \phi$
      desugars to $\neg \X \neg \phi$. By Lemma~\ref{lem:commnegnext},
      we have $\END \vee \X \neg \neg \phi$. If $\END$ holds, we are
      done immediately by \rn{Taut}. Otherwise, we have $\X \neg \neg
      \phi$, which gives us $\X \phi$ by \rn{Taut}, as well.

    From right to left, suppose $\vdash \X \phi \vee \END$. By the
      law of the excluded middle, we have $\END \vee \neg \END$
      (\rn{Taut}). If $\END$ holds, then we are done immediately by \rn{EndNextContra}.
      If $\X \phi \wedge \neg \END$ holds, then we can show $\vdash
      \WX \phi$ by desugaring to $\X \neg \phi \Implies
      \bot$. Suppose for a contradiction that $\X \neg \phi$. We
      have $\X \phi$ and $\X \neg \phi$, so $\X \bot$. But from
      $\neg \END$, we have $\X \top$---and by
      Lemma~\ref{lem:nextcontra} we have a contradiction.
  \end{proof}
\end{lemma}

\begin{lemma}[Weak next negation before the end]
  \label{lem:wknextnegend}
  \[ \vdash \neg \END \wedge \WX \neg \phi \Implies \neg \WX \phi \]
  \begin{proof}
    We have $\neg \END \wedge \WX \neg \phi$.
    By unrolling syntax, we have $\neg \END \wedge \neg \X \neg \neg \phi$.
    By \rn{Taut}, we have $\neg \END \wedge \neg \X \phi$.
    By Lemma~\ref{lem:nextwknext}, $\X \phi \Implies \WX \phi$, so we have $\neg \WX \phi$.
  \end{proof}
\end{lemma}

\begin{lemma}[Weak next distributes over conjunction]
  \label{lem:wknextdistconj}
  \[\vdash \WX(\phi \wedge \psi) \Iff \WX \phi \wedge \WX \psi.\]
  \begin{proof}
    From $\WX(\phi \Implies \psi)$, we have $\WX( \neg(\phi \Implies \neg \psi))$ by \rn{Taut}.
    By \rn{Taut} again, we have $\END \vee \neg\END$; by the definition of $\WX$, we can refactor our formula into
    $\WX(\neg(\phi \Implies \neg \psi)) \wedge \END \vee \neg \X(\phi \Implies \neg \psi) \wedge \neg \END$, also by \rn{Taut}.
    By \rn{EndNextContra}, we have $\END \Implies \neg \X \neg \neg
    (\phi \Implies \neg \psi)$, so we can eliminate the left-hand
    disjunct by \rn{Taut}, to find
    $\END \vee \neg \X(\phi \Implies \neg \psi) \wedge \neg \END$.
    By Lemma~\ref{lem:nextwknext}, we can weaken our next modality to
    find $\END \vee \neg \WX(\phi \Implies \neg \psi) \wedge
    \neg \END$.
    By \rn{WkNextDistr}, we distribute the modality across the
    implication and we have $\END \vee \neg (\WX \phi \Implies \WX
    \neg \psi) \wedge \neg \END$.
    Using Lemma~\ref{lem:wknextnegend}, we can change $\WX \neg \psi$
    to $\neg \WX \psi$, because we have $\neg \END$ in that disjunct; we
    now have $\END \vee \neg (\WX \phi \Implies \neg \WX \psi) \wedge
    \neg \END$.
    By \rn{Taut}, we can rearrange the implication to find $\END \vee
    (\WX \phi \wedge \WX \psi) \wedge \neg \END$.
    By \rn{EndNextContra}, we can introduce $\WX \phi$ and $\WX \psi$
    on the left-hand disjunct, to find $\WX \phi \wedge \WX \psi
    \wedge \END \vee \WX \phi \wedge \WX \psi \wedge\neg \END$.
    Finally, \rn{Taut} allows us to rearrange our term to $\WX \phi
    \wedge \WX \psi \wedge (\END \vee \neg\END)$, where the rightmost
    conjunct falls out and we find $\WX \phi \wedge \WX \psi$.
  \end{proof}
\end{lemma}

\begin{lemma}[Weak next negation]
  \label{lem:wknextneg}
  $\vdash \neg \WX \phi \Implies \WX \neg \phi$
  \begin{proof}
    By definition, $\neg \WX \phi$ is equivalent to $\neg \neg \X \neg \phi$.
    By \rn{Taut}, we eliminate the double negation to find $\X \neg \phi$.
    By Lemma~\ref{lem:commnegnext} from right to left, we have $\neg \X \phi$.
    By \rn{Taut} , we reintroduce an inner double negation, to find
    $\neg \X \neg \neg \phi$.
    By definition, we have $\WX \neg \phi$.
  \end{proof}
\end{lemma}

\begin{lemma}[Always unrolling]
  \label{lem:alwaysunroll}
  $\vdash \A \phi \Iff \phi \wedge \WX \A \phi$
  \begin{proof}
    Desugaring, we must show $\vdash (\phi \W \bot) \Iff \phi \wedge
    \WX (\phi \W \bot)$. By \rn{WkUntilUnroll}, $\vdash (\phi \W \bot)
    \Iff \bot \vee (\phi \wedge \WX (\phi \W \bot))$. By \rn{Taut} we
    can eliminate the $\bot$ case of the disjunction on the right.
  \end{proof}
\end{lemma}

\begin{lemma}[Necessitation]
  \label{lem:boxstep}
  If $\vdash \phi$ then $\vdash \A \phi$.
  \begin{proof}
    We apply \rn{Induction} with $\phi = \top$ and $\psi = \phi$ to
    show that $\vdash \top \Implies \A \phi$, i.e., $\vdash \A \phi$
    by \rn{Taut}.
    We must prove both premises: $\vdash \top \Implies \phi$ and
    $\vdash \top \Implies \WX \top$.
    Since we've assumed $\vdash \phi$, we have the first premise by \rn{Taut}.
    By \rn{Taut} and \rn{WkNextStep}, we have $\vdash \WX \top$, and
    so $\vdash \top \Implies \WX \top$ by \rn{Taut}.
  \end{proof}
\end{lemma}

\subsection{Completeness}
\label{sec:ltlf_completeness}

To show deductive completeness for \LTLf, we must find that if
$\models \phi$ then $\vdash \phi$. To do so we construct a graph
that does two things at once: first, paths from the root of the graph
to a terminal state correspond to Kripke structures which $\phi$
satisfies; second, consistency properties in the graph relate to the
provability of the underlying formula $\phi$.

\begin{figure}[tp]

\pgfdeclarelayer{bg}
\pgfdeclarelayer{fg}
\pgfsetlayers{bg,main,fg}

\definecolor{pnpcolor}{RGB}{253,174,97}
\definecolor{graphcolor}{RGB}{43,131,186}
\definecolor{terminalcolor}{RGB}{171,221,164}
\definecolor{completenesscolor}{RGB}{215,25,28}

\tikzstyle{lem}=[draw,rectangle,rounded corners,align=center]
\tikzstyle{lemgrp}=[lem,fill=yellow!20,draw=black!50,dashed]
\tikzstyle{pnps}=[lem,pnpcolor,fill=pnpcolor,text=white]
\tikzstyle{proofgraph}=[lem,graphcolor,fill=graphcolor,text=white]
\tikzstyle{terminalpaths}=[lem,terminalcolor,fill=terminalcolor,text=white]
\tikzstyle{completeness}=[lem,completenesscolor,fill=completenesscolor,text=white]
\tikzstyle{imp}=[->,thick,double,rounded corners]

\centering
\begin{tikzpicture}[thick, double, node distance=3cm]
  \node [pnps] (pnpproperties) {PNP \\ properties \\ Lemma~\ref{lem:pnpproperties}};

  \node [pnps] (transprovable) [right of=pnpproperties,node distance=3.25cm] {Provable \\ Lemma~\ref{lem:transprovable}};
  \node (transitions) [above right of=transprovable,node distance=1.75cm] {Transitions};
  \node [pnps] (transconsistent) [below right of=transitions, node distance=1.75cm] {Consistent \\ Lemma~\ref{lem:transconsistent}};
  \begin{pgfonlayer}{bg}
    \path (transprovable.west |- transitions.north)+(-0.3,0.3) node (tl) {};
    \path (transconsistent.east |- transconsistent.south)+(0.3,-0.3) node (br) {};
    \path [lemgrp] (tl) rectangle (br);
  \end{pgfonlayer}

  \node [pnps] (inconsistentcompletions) [below of=pnpproperties, node distance=2.25cm] {Inconsistent \\ $\Implies$ \\ $\nexists$ completions \\ Lemma~\ref{lem:inconsistentcompletions}};

  \node [pnps] at (transconsistent |- inconsistentcompletions) (completionsprovable) {Completions \\ provable \\ Lemma~\ref{lem:completionsprovable}};

  \node (assignments) [right of=transitions,node distance=4.25cm] {Assignments};
  \node [pnps] (closureprovable) at (assignments |- transconsistent) {Provable \\ Lemma~\ref{lem:closureprovable}};
  \node [pnps] (assigncomps) at (assignments |- inconsistentcompletions) {Consistent \\ $\Implies$ \\ Completion \\ Lemma~\ref{lem:assigncomps}};
  \begin{pgfonlayer}{bg}
    \path (assigncomps.west |- assignments.north)+(-0.3,0.3) node (tl) {};
    \path (assigncomps.east |- assigncomps.south)+(0.3,-0.3) node (br) {};
    \path [lemgrp] (tl) rectangle (br);
  \end{pgfonlayer}

  \draw[imp] (transprovable) -- (transconsistent);
  \draw[imp] (closureprovable) -| ([xshift=-0.45cm,yshift=0.4cm] assigncomps.west) -- ([yshift=0.4cm] completionsprovable.east);
  \draw[imp] (assigncomps) -- (completionsprovable);

  \node [proofgraph] (endinvariant) [below of=inconsistentcompletions,node distance=2.25cm] {Invariance \\ of $\END$ \\ Lemma~\ref{lem:endinvariant}};
  \node [proofgraph] (nodesconsistent) [below of=endinvariant,node distance=2.25cm] {Graph \\ consistent \\ and complete \\ Lemma~\ref{lem:nodesconsistent}};
  \node [proofgraph] (succprovable) at (transprovable |- endinvariant) {Transition \\ implication \\ Lemma~\ref{lem:succprovable}};

  \node [terminalpaths] (graphoperators) at (succprovable |- nodesconsistent) {Proof graphs \\ are models \\ Lemma~\ref{lem:graphoperators}};

  \draw[imp] (pnpproperties) -- ([xshift=-0.5cm] transprovable.west) |- (succprovable.west);
  \draw[imp] (transprovable) -- (succprovable);
  \draw[imp] (transconsistent) |- ([xshift=0.4cm,yshift=-0.75cm] transprovable.south) -- ([xshift=0.4cm] succprovable.north);
  \draw[imp] (completionsprovable) -| ([xshift=0.8cm] succprovable.north);

  \draw[imp] (inconsistentcompletions) -- (endinvariant);
  \draw[imp] (endinvariant) -- (nodesconsistent);

  \node [terminalpaths] (terminalpaths) at (assigncomps |- graphoperators) {Paths \\ Corollary~\ref{cor:terminalpaths}};
  \node (terminality) [above left of=terminalpaths, node distance=1.75cm] {Terminal objects exist};
  \node [terminalpaths] (terminalnodes) [below left of=terminality,node distance=1.75cm] {Nodes \\ Lemma~\ref{lem:terminalnodes}};
  \begin{pgfonlayer}{bg}
    \path (terminalnodes.west |- terminality.north)+(-0.3,0.3) node (tl) {};
    \path (terminalpaths.east |- terminalpaths.south)+(0.3,-0.3) node (br) {};
    \path [lemgrp] (tl) rectangle (br);
  \end{pgfonlayer}

  \draw[imp] (terminalnodes) -- (terminalpaths);
  \draw[imp] (succprovable) -| ([xshift=-0.5cm] terminalnodes.west) -- (terminalnodes);

  \draw[imp] ([yshift=-0.2cm] pnpproperties.east) -- ([xshift=-0.7cm,yshift=-0.2cm] transprovable.west) |- ([yshift=0.5cm] graphoperators.north) -- (graphoperators.north);
  \draw[imp] (nodesconsistent) -- (graphoperators);

  \node [completeness] (ltlfcompletenessctx) [below of=terminalpaths,node distance=3.25cm] {w/Context \\ Corollary~\ref{cor:ltlfcompletenessctx}};
  \node (completeness) [above left of=ltlfcompletenessctx,node distance=2cm] {Completeness};
  \node [completeness] (ltlfcompleteness) [below left of=completeness,node distance=2cm] {Formulae \\ Theorem~\ref{thm:ltlfcompleteness}};
  \begin{pgfonlayer}{bg}
    \path (ltlfcompleteness.west |- completeness.north)+(-0.3,0.3) node (tl) {};
    \path (ltlfcompletenessctx.east |- ltlfcompletenessctx.south)+(0.3,-0.3) node (br) {};
    \path [lemgrp] (tl) rectangle (br);
  \end{pgfonlayer}

  \node [completeness] (ltlfsatisfiability) at ([xshift=-4.35cm] ltlfcompleteness) {Satisfiability \\ Theorem~\ref{thm:ltlfsatisfiability}};

  \draw[imp] (nodesconsistent.south) |- ([xshift=-0.4cm,yshift=0.75cm] ltlfsatisfiability.north) -- ([xshift=-0.4cm] ltlfsatisfiability.north);
  \draw[imp] (terminalpaths.south) |- ([xshift=0.4cm,yshift=-0.3cm] graphoperators.south) |- ([xshift=0.4cm,yshift=0.5cm] ltlfsatisfiability.north) -- ([xshift=0.4cm] ltlfsatisfiability.north);
  \draw[imp] (graphoperators.south) |- ([yshift=0.8cm] ltlfsatisfiability.north) -- (ltlfsatisfiability.north);
  \draw[imp] (pnpproperties.west) -| ([xshift=-0.2cm] nodesconsistent.west) |- (ltlfsatisfiability.west);

  \draw[imp] (ltlfsatisfiability) -- (ltlfcompleteness);
  \draw[imp] (ltlfcompleteness) -- (ltlfcompletenessctx);
\end{tikzpicture}

Proof sections, by color: \\[1em]

\begin{tikzpicture}[start chain, every node/.style={on chain,align=center}]
\node [pnps] (pnps) {PNPs \\ (Nodes)};
\node [proofgraph] (proofgraph) {Proof \\ graph};
\node [terminalpaths] (terminal) {Paths and \\ models};
\node [completeness] (completeness) {Completeness};
\end{tikzpicture}

\caption{Completeness for \LTLf}
\label{fig:proofstructure}
\end{figure}

Our construction follows the standard Henkin-Hasenjaeger-style least-fixed point approach found in 
Kr\"oger and Merz's book~\cite{KrogerMerz_LTL_2008}: we
construct a graph whose nodes assign truth values to each subformula
of our formula of interest, $\phi$, by putting each subformula in
either the true, ``positive'' set or in the false, ``negative'' set.
To show that $\models \phi$ implies $\vdash \phi$, we build a graph for $\neg \phi$ that guarantees that $\neg \phi$ is inconsistent, i.e., $\vdash \neg \neg \phi$, and so $\vdash \phi$ via double-negation elimination (since \LTLf's propositional core is classical). Our
proof itself is classical, using the law of the excluded middle to
define the proof graph (\comps) and prove some of its properties
(Lemma~\ref{lem:graphoperators}).

What about the `$f$' in \LTLf? Nothing described so far differs in any way from
the Henkin-Hasenjaeger graph approach used by Kr\"oger and
Merz~\cite{KrogerMerz_LTL_2008}. Kr\"oger and Merz's graphs were always finite,
but their notion of satisfying paths forces paths to be infinite. We restrict
our attention to terminating paths: paths where not only is our formula of
interest satisfied, but so is $\E \END$. To ensure such paths exist,
we \emph{inject} $\E \END$ into the root node of the graph.

The proof follows the following structure (Figure~\ref{fig:proofstructure}): we define the nodes of the
graph (Definition~\ref{def:pnp}); we define the edge relation on the
graph (Figure~\ref{fig:stepclosure}) and show that it maps
appropriately to time steps in the proof theory
(Lemma~\ref{lem:completionsprovable} finds a consistent successor;
Lemma~\ref{lem:transprovable} shows the successor is a state in our
graph); we show that the graph structure results in a finite structure
with appropriate consistency properties
(Lemma~\ref{lem:succprovable}); we define which paths in the graph
represent our Kripke structure of interest
(Lemma~\ref{lem:graphoperators} shows that our graph's transitions
correspond to the semantics; Lemma~\ref{cor:terminalpaths} guarantees
that we have appropriate finite models). The final proof comes in two
parts: we show that consistent graphs correspond to satisfiable
formulae (Theorem~\ref{thm:ltlfsatisfiability}), which we then use to
show completeness (Theorem~\ref{thm:ltlfcompleteness}).

\begin{definition}[PNP]
  \label{def:pnp}
  A \emph{positive-negative pair} (PNP) $\P$ is a pair of finite sets of
  formulae $(\pos(\P), ~ \negs(\P))$. We refer to the collected
  formulas of $\P$ as $\F_\P = \pos(\P) \cup \negs(\P)$; we call the
  set of all PNPs $\PNP$.

  We write the \emph{literal interpretation} of a PNP $\P$ as: \[\Phat =
  \bigwedge_{\mathclap{\phi \in \pos(\P)}} \phi \hspace{.7em} \wedge \hspace{.7em} \bigwedge_{\mathclap{~\psi \in \negs(\P)}} \neg \psi.\]

  We say $\P$ is \emph{inconsistent} if $\vdash \neg \Phat$;
  conversely, $\P$ is \emph{consistent} when it is not the case that
  $\vdash \neg \Phat$, i.e., $\nvdash \neg\Phat$.
\end{definition}

Positive-negative pairs are the nodes of our proof graph---
each node is a collection of formulae that hold (or
not) in a given moment in time.
Before we can even begin constructing
the graph, we show that they adequately characterize a moment in time:
that is, they are without contradiction, can be `saturated' with all
of the formulae of interest, and respect the general rules of our
logic.
Readers may be familiar with `atoms', but PNPs are themselves not
atoms; \emph{complete} PNPs are more or less atoms
(Figure~\ref{fig:stepclosure}).

\begin{lemma}[PNP properties]
  \label{lem:pnpproperties}

  For all consistent PNPs $\P$:
  \begin{enumerate}
  \item $\pos(\P) \cap \negs(\P) = \emptyset$;
  \item For all $\phi$, either $(\pos(\P) \cup \{ \phi \}, ~ \negs(P))$
    or $(\pos(\P), ~ \negs(P) \cup \{ \phi \})$ is consistent;
  \item $\bot \not\in \pos(\P)$;
  \item if $\{ \phi, \psi, \phi \Implies \psi \} \subseteq \F_\P$,
    then $\phi \Implies \psi \in \pos(\P)$ iff $\phi \in \negs(\P)$ or $\psi \in \pos(P)$;
  \item if $\vdash \phi \Implies \psi$ and $\phi \in \pos(\P)$ and
    $\psi \in \F_\P$, then $\psi \in \pos(\P)$.
  \end{enumerate}
  \begin{proof}
    Let a given PNP $\P$ be consistent.
    We show each case by reasoning based on whether each formula is assigned to
    the positive or the negative set of $\P$, deriving contradictions as
    appropriate.
    \begin{enumerate}
    
    \item Suppose for a contradiction that
      $\phi \in \pos(\P) \cap \negs(\P)$. We have $\vdash \neg
      (\phi \wedge \neg\phi)$ by \rn{Taut}, but
      $\vdash \Phat \Implies \phi \wedge \neg \phi$, and so $\vdash \neg \Phat$
      by \rn{Taut}---making $\P$ inconsistent, a contradiction.

    \item If $\phi \in \pos(\P)$ or $\phi \in \negs(\P)$ already, we are done;
      we already know by (1) that $\phi \not\in \pos(\P) \cap \negs(\P)$.
      So $\phi$ does not already occur in $\P$. Suppose for a contradiction
      that adding $\phi$ to either set is inconsistent, i.e.  both $\vdash \neg
      (\Phat \wedge \phi)$ and $\vdash \neg (\Phat \wedge \neg
      phi)$. By \rn{Taut}, that would imply that $\vdash \neg \Phat \wedge
      (\phi \vee \neg \phi)$, which is the same as simply
      $\vdash \neg \Phat$---a contradiction.
      
    \item Suppose for a contradiction that $\bot \in \pos(\P)$; by \rn{Taut} we
      have $\vdash \Phat \Implies \bot$, which is syntactic sugar for
      $\vdash \neg \Phat$---a contradiction.
      
    \item Suppose $\{ \phi, \psi, \phi \Implies \psi \} \subseteq \F_\P$.
        If $\phi \Implies \psi \in \pos(\P))$, we must show that
          $\phi \in \pos(\P)$ or that $\psi \in \negs(\P)$. Suppose for a
          contradiction that neither is in the appropriate set; we then have
          $\vdash \Phat \Implies
          (\phi \Implies \psi) \wedge \phi \wedge \neg \psi$; by \rn{Taut}, we
          can then conclude $\vdash \neg \Phat$---a contradiction.

        If, on the other hand, $\phi \in \negs(\P)$ or $\psi \in \pos(\P)$, we must show that
          $\phi \Implies \psi \in \pos(\P)$. Suppose for a contradiction that
          its not the case that $\phi \Implies \psi \in \pos(\P)$. Since
          $\phi \Implies \psi \in \F_\P$, then
          $\phi \Implies \psi \in \negs(\P)$. We have either
          $\vdash \Phat \Implies \neg (\phi \Implies \psi) \wedge \neg \phi$ or
          $\vdash \Phat \Implies \neg
          (\phi \Implies \psi) \wedge \psi$. By \rn{Taut}, we can convert $\neg
          (\phi \Implies \psi)$ into $\phi \wedge \neg \psi$---and either way we
          can find by \rn{Taut} that $\vdash \neg \Phat$, a
          contradiction.

    \item Suppose $\vdash \phi \Implies \psi$ and $\psi \in \pos(\P)$ with
      $\phi \in \F_\P$. We must show that $\phi \in \pos(\P)$. Suppose for a
      contradiction that $\phi \in \negs(\P)$. We then have
      $\vdash \Phat \Implies (\phi \Implies \psi) \wedge \psi \wedge \neg \phi$;
      by \rn{Taut}, we can then find $\vdash \neg \Phat$, which is a
      contradiction. \qedhere

  \end{enumerate}
\end{proof}
\end{lemma}

\begin{figure}[t]
  \hdr{Transition functions}{\fbox{$\sigma^{\bullet}_i : \PNP \rightarrow 2^{\LTLf}$} \quad \fbox{$\sigma : \PNP \rightarrow \PNP\vphantom{2^{\LTLf}\sigma^{\bullet}_i}$}}

  \[ \begin{array}{rcl}
    \sigma^+_1(\P) &=& \{ \phi         \mid \X \phi \in \pos(\P) \} \\[.5em]
    \sigma^+_2(\P) &=& \{ \phi \W \psi \mid \phi \W \psi \in \pos(\P), \psi \in \negs(\P) \} \\[.5em]
    \sigma^-_3(\P) &=& \{ \phi         \mid \X \phi \in \negs(\P) \} \\[.5em]
    \sigma^-_4(\P) &=& \{ \phi \W \psi \mid \phi \W \psi \in \negs(\P), ~ \phi \in \pos(\P) \} \\[1em]

    \sigma(\P) &=& (\sigma^+_1(\P) \cup \sigma^+_2(\P), ~ \sigma^-_3(\P) \cup \sigma^-_4(\P))
  \end{array} \]


  \[ \begin{array}{rcl@{\quad}rcl}
    \tau(v)                  &=& \{ v \} &
    \tau(\bot)               &=& \{ \bot \} \\
    \tau(\phi \Implies \psi) &=& \{ \phi \Implies \psi \} \cup \tau(\phi) \cup \tau(\psi) &
    \tau(\X \phi)            &=& \{ \X \phi \} \\
    \tau(\phi \W \psi)       &=& \{ \phi \W \psi \} \cup \tau(\phi) \cup \tau(\psi) \\[1em]

    \tau(\F) &=& \bigcup_{\phi \in \F} \tau(\phi) &
    \tau(\P) &=& \tau(\F_\P)
  \end{array} \]

  \hdr{Extensions, completions, and possible assignments}{
       \fbox{$\mathord{\preceq} \subseteq \PNP \times \PNP\vphantom{2^{\PNP}}$} 
       \quad \fbox{$\comps : \PNP \rightarrow 2^\PNP$} 
       \quad \fbox{$\assigns : 2^{\LTLf} \rightarrow 2^{\PNP}$}}

  \[ \begin{array}{@{}c@{}}
    \P \preceq \Q \text{ iff } \pos(\P) \subseteq \pos(\Q) \text{ and } \negs(\P) \subseteq \negs(\Q) \\[.5em]
    \assigns(\F) = \{ \P \mid \F_\P = \tau(\F) \}
    \\[.5em]
    \comps(\P) = \{ \Q \mid \F_\Q = \tau(\P), ~ \P \preceq \Q, ~ \Q \text{ consistent} \}
  \end{array} \]

  \caption{Step and closure functions; extensions and completions}
  \label{fig:stepclosure}
\end{figure}

Our goal is to generate successors states to build a graph of PNPs; to
do so, we define two functions: a step function $\sigma$ and a closure
function $\tau$ (Figure~\ref{fig:stepclosure}).
The closure function $\tau$ takes a PNP and produces all of its
subterms that are relevant for the current state, i.e., it doesn't go
under the next modality.
Our closure $\tau$ is slightly smaller than the commonly seen
Fischer-Ladner closure~\cite{PDL}: we don't include every possible negation and
we stop when we reach a next modality.
We write $\P \preceq \Q$ (read ``\P is extended by \Q'' or ``\Q extends \P'') when $\Q$'s
positive and negative sets subsume $\P$'s
(Figure~\ref{fig:stepclosure}). We say $\P$ is complete when $\F_\P =
\tau(\P)$. We say a complete PNP $\Q$ is a \emph{completion} of $\P$
when $\P \preceq \Q$ and $\Q$ is consistent and complete. 
We define the set of all consistent completions of a given PNP $\P$
as $\comps(\P)$.
The step function
$\sigma$ takes a PNP and generates those formulae which must hold in
the next step, thereby characterizing the transitions in our
graph.
The set of completions, $\comps$, is not a constructive set, since we have (as
yet) no way to determine whether a given PNP is consistent or not. 

First, we show that each PNP implies its successor (Lemma~\ref{lem:transprovable}); next, consistent PNPs produce consistent successors (Lemma~\ref{lem:transconsistent}).

\begin{lemma}[Transitions are provable]
  \label{lem:transprovable}
  For all $\P \in \PNP$, we have $\vdash \Phat \Implies \WX
  \widehat{\sigma(\P)}$.
  \begin{proof}
    Unfolding the definition of $\Phat$, we must show
    \[
    \vdash
      \left[ \hspace{1.05em}
      \bigwedge_{\mathclap{\phi \in \pos(\P)}} \phi
      \hspace{.7em} \wedge \hspace{.7em}
      \bigwedge_{\mathclap{\psi \in \negs(\P)}} \psi
      \hspace{.4em} \right]
      \Implies
      \WX \left[ \hspace{1.55em}
      \bigwedge_{\mathclap{\phi \in \pos(\sigma(\P))}} \phi
      \hspace{.9em} \wedge \hspace{.9em}
      \bigwedge_{\mathclap{\psi \in \negs(\sigma(\P))}} \psi
      \hspace{.9em} \right].
    \]
        By cases on the
    clauses of $\sigma$, we show that $\Phat$ implies each of the parts of
    $\widehat{\sigma(\P)}$, tying the cases together by \rn{Taut} and
    Lemma~\ref{lem:wknextdistconj}:
    
    \begin{enumerate}[align=left]
    
    \item[($\sigma^+_1$)] Suppose $\phi \in \sigma^+_1(\P)$ because
      $\X \phi \in \P$. We have $\vdash \Phat \Implies \WX \phi$ because
      $\X \phi \Implies \WX \phi$ by Lemma~\ref{lem:nextwknext}.
      
    \item[($\sigma^+_2$)] Suppose $\phi \W \psi \in \sigma^+_2(\P)$ because
      $\vdash \phi \W \psi \in \pos(\P)$ and $\psi \in \negs(\P)$. Then,
      $\vdash \Phat \Implies \neg \psi \wedge \phi \W \psi$. We have
      $\Phat \Implies \WX \phi \W \psi$ by \rn{WkUntilUnroll} and \rn{Taut}.
      
    \item[($\sigma^-_3$)] Suppose $\phi \in \sigma^-_3(\P)$ because
      $\X \phi \in \negs(\P)$. We have $\vdash \Phat \Implies \WX \neg \phi$
      because $\neg \X \phi \Implies \WX \neg \phi$ by
      Lemmas~\ref{lem:commnegnext} and~\ref{lem:nextwknext}.
    \item[($\sigma^-_4$)] Suppose $\phi \W \psi \in \sigma^-_4(\P)$
      because $\phi \W \psi \in \negs(\P)$ and $\phi \in \pos(\P)$. We
      have:
      \[ \begin{array}{r@{~~\vdash {}}lrr}
      & \Phat \Implies \neg (\phi \W \psi) \wedge \phi && \\
      \text{iff} & \Phat \Implies \neg (\psi \vee (\phi \wedge \WX(\phi \W \psi))) \wedge \phi & \rn{WkUntilUnroll} & \\
      \text{iff} & \Phat \Implies \neg \psi \wedge \neg (\phi \wedge \WX(\phi \W \psi)) \wedge \phi & \rn{Taut} & \\
      \text{iff} & \Phat \Implies \neg \psi \wedge (\neg \phi \vee \neg \WX(\phi \W \psi)) \wedge \phi & \rn{Taut} & \\
      \text{iff} & \Phat \Implies \neg \psi \wedge \neg \WX(\phi \W \psi) & \rn{Taut} & \\
      \text{implies} & \Phat \Implies \neg \psi \wedge \WX \neg (\phi \W \psi) \wedge \phi & Lemma~\ref{lem:wknextneg} & \\
      \text{iff} & \Phat \Implies \WX \neg (\phi \W \psi) & \rn{Taut} & \qedhere \\
      \end{array} \] 
    \end{enumerate}
  \end{proof}
\end{lemma}

\smallskip

\begin{lemma}[Transitions are consistent]
  \label{lem:transconsistent}
  For all consistent PNPs $\P$, if $\vdash \Phat
  \Implies \neg \END$ then $\sigma(\P)$ is consistent.
  \begin{proof}

    Let $\P$ be a consistent PNP such that $\vdash \Phat \Implies \neg \END$.
    Assume for the sake of contradiction, that
    $\vdash \neg \widehat{\sigma(\P)}$. Then we can write, by \rn{WkNext},
    $\vdash \WX \neg \widehat{\sigma(\P)}$, which is equivalent to
    $ \vdash \neg \X \widehat{\sigma(\P)}$.
    By Lemma~\ref{lem:transprovable}, $\vdash \Phat \Implies \WX
    (\widehat{\sigma(\P)}) \wedge \neg \END$, which by
    Lemma~\ref{lem:nextwknext} and \rn{Taut} gives $\vdash \Phat \Implies \X
    (\widehat{\sigma(\P)})$.
    Now we can derive $\vdash \Phat \Implies \bot$, or equivalently
    $\vdash \neg \Phat$---a contradiction.
    Therefore $\sigma(\P)$ is consistent.
  \end{proof}
\end{lemma}

Having established the fundamental properties of our successors, we
must \emph{complete} them: each PNP state needs to be `saturated' to
include all formulae of interest from the previous state.
A consistent PNP has many such possible completions---and we
prove as much below---but we first observe that an inconsistent PNP
has no completions.

\begin{lemma}[Inconsistent PNPs have no completions]
  \label{lem:inconsistentcompletions}
  If a PNP $\P$ is inconsistent, then $\comps(\P) = \emptyset$.
  \begin{proof}
    Let $\P$ be given; suppose for a contradiction that there exists
    $\Q \in \comps(\P)$, i.e, $\F_\Q = \tau(\P)$ and $\P \preceq \Q$
    and $\Q$ is consistent. We have $\vdash \Qhat \Implies \Phat$ by
    \rn{Taut}, because $\Q$ is an extension of $\P$. But we know that
    $\vdash \neg \Phat$, so it must be the case that $\vdash \neg
    \Qhat$---which would mean that $\Q$ was inconsistent, a
    contradiction.
  \end{proof}
\end{lemma}

In order to fully define our graph, we must show that not only are
successors of PNPs provable, so are their completions. We do so in two
steps: first, we show that there is always \emph{some} provable
assignment of propositions in each set of formulas; next,
conditionally provable assignments are in fact completions.

\begin{lemma}[Assignments are provable]
  \label{lem:closureprovable}
  $\vdash \bigvee_{\P \in \assigns(\F)} \Phat$
  \begin{proof}
    By induction on the size of $\F$.
    When $|\F| = 0$, We have $\vdash \top$ by \rn{Taut}.
     When $|\F| = n+1$, let $\phi \in \F$ be a maximal formula, i.e.,
      $\phi \not\in \tau(\F - \{ \phi \})$.%
      We have $\assigns(\F) = \{ \P \mid \P' \in \assigns(\F'), ~ \F_\P
      = \F_{\P'} \cup \tau(\phi) \}$, i.e., each formula in $\tau(\phi)$ not
      already assigned in $\P'$ is put in either the positive or negative set of
      $\P$. That is, we take each formula in $\P'$ and conjoin
      $\psi \vee \neg\psi$ for each $\psi \in \tau(\phi)$.
      We know by the IH that
      $\vdash \bigvee_{\P' \in \assigns(\F')} \Phat'$, so by \rn{Taut} we have
      $\vdash \bigvee_{\P \in \assigns(\F)} \Phat$.
    \end{proof}
\end{lemma}

\smallskip

\begin{lemma}[Consistent assignments are completions] ~ \\
  \label{lem:assigncomps}
  For all consistent PNPs $\P$ and for all $\Q \in \assigns(\Phat)$, if
  $\vdash \Phat \Implies \Qhat$ then $\Q \in \comps(\P)$.
  \begin{proof}
    Let $\P$ and $\Q \in \assigns(\Phat)$ be given such that $\vdash
    \Phat \Implies \Qhat$.

    Suppose for a contradiction that $\Q \not\in \comps(\P)$. It must
    be the case that either $\Q$ does not extend $\P$ or $\Q$ is
    inconsistent---\iffull{}we show that \fi{}both cases are contradictory.

    If $\P \npreceq \Q$, then there exists some formula $\phi$ such
    that $\phi \in \pos(\P)$ and $\phi \in \negs(\Q)$ or vice versa (since $\Q \in \assigns(\Phat)$, every formula must be accounted for).
      Then, $\vdash \Phat \Implies \phi$ and
      $\vdash \Qhat \Implies \neg \phi$. Then
      $\vdash \Phat \wedge \Qhat \Implies \phi \wedge \neg\phi$, which
      by \rn{Taut} means $\vdash \neg (\Phat \wedge \Qhat)$, or equivalently,
      that $\vdash \Phat \Implies \neg \Qhat$. When combined with the assumption
      that $\vdash \Phat \Implies \Qhat$, via \rn{Taut}, we can derive
      $\vdash \neg \Phat$---a contradiction with $\P$'s consistency.
      
    If, on the other hand $\Q$ is inconsistent, then we can see from
      $\vdash \Phat \Implies \Qhat$ that $\vdash \neg \Qhat$---and
      by \rn{Taut}, it must be that $\vdash \neg \Phat$, which
      contradicts $\P$'s consistency. \end{proof}
\end{lemma}

Combining the last two proofs we find that consistent completions are
provable.

\begin{lemma}[Consistent completions are provable] ~ \\
  \label{lem:completionsprovable}
  For all consistent PNPs $\P$, we have $\vdash \Phat
  \Implies \bigvee_{\Q \in \comps(\P)}
  \Qhat$.
  \begin{proof}
    By Lemma~\ref{lem:closureprovable}, we have $\vdash \bigvee_{\Q
      \in \assigns(\Phat)} \Qhat$. By \rn{Taut}, we have $\vdash \Phat
    \Implies \bigvee_{\Q \in \assigns(\Phat)} \Qhat$. By
    Lemma~\ref{lem:assigncomps}, we know that we only need to keep
    those $\Q \in \assigns(\Phat)$ which are also in $\comps(\P)$, and
    so we have $\vdash \Phat \Implies \bigvee_{\Q \in \comps(\P)}
    \Qhat$ as desired.
  \end{proof}
\end{lemma}
Having established the fundamental properties of consistent completions, we set
about defining proof graphs, the structure on which we build our proof. Starting
from a PNP formed from a given formula, we can construct a graph where nodes are
PNPs and a node $\P$'s successors are consistent completions of $\sigma(\P)$.

\begin{definition}[Proof graphs]
  \label{def:proofgraph} For a consistent and complete PNP
  $\P$\iffull{} (i.e., where $\F_\P = \tau(\P)$ and it is not the case
  that $\vdash \neg \Phat$)\fi, we define a \emph{proof graph} $\G_\P$
  as follows: (a) \P is the \emph{root} of $\G_\P$; (b) \P has an edge
  to the root of $\G_\Q$ for each $\Q \in \comps(\sigma(\P))$.
%
\end{definition}
Since $\P$ is composed of a finite number of formulae, the set of all
  subsets of $\tau(\P)$ is finite, as are any assignments of those
  subsets to PNPs. Hence the number of nodes in the proof graph must
  be finite.\footnote{Confusingly, Kr\"oger and
  Merz~\cite{KrogerMerz_LTL_2008} call this graph an ``infinite tree''
  in their proof of completeness for potentially infinite LTL, even
  though it turns out to be finite in that setting, as well.}

Our innovation in adapting the completeness proof to finite time is 
\emph{finiteness injection}, where we make sure that $\E \END$ is in the 
positive set of the root of the proof graph.
After injecting finiteness, every node of the proof graph  either has
$\END$ in its positive set (and no successors) or all of its successors have 
$\E \END$ in their positive set.

Every lemma we prove, from here to the final completeness result, has
some premise concerning the end of time: by only working with
PNPs with $\E \END$ in the positive set, we guarantee that time
eventually ends.

\begin{lemma}[\END injection is invariant]
  \label{lem:endinvariant}
  If \P is a consistent and complete PNP with $\E \END \in \pos(\P)$, then either:
  \begin{itemize}
  \item $\END \in \pos(\P)$ and $\P$ has no successors (i.e., $\comps(\sigma(\P)) = \emptyset$), or
  \item $\END \in \negs(\P)$ and for all $\Q \in \comps(\sigma(\P))$, we have $\E \END \in \pos(\Q)$.
  \end{itemize}
  \begin{proof}
    Recall that $\E \END$ desugars to $\neg (\neg \neg \X \top \W
    \bot)$. Since $\P$ is complete, we know that $\END \in \F_\P$.
    If $\END$ (i.e., $\neg \X \top$) is in
      $\pos(\P)$ and $\P$ is consistent, it must be the case that $\X
      \top \in \negs(\P)$ by Lemma~\ref{lem:pnpproperties}. We
      therefore have that $\top \in \sigma^-_3(\P)$, so $\vdash
      \widehat{\sigma(\P)} \Implies \neg \top$, i.e., $\sigma(\P)$ is
      inconsistent---and therefore $\comps(\sigma(\P)) = \emptyset$,
      because there are no consistent completions of an inconsistent
      PNP (Lemma~\ref{lem:inconsistentcompletions}).
    If, on the other hand, $\END$ (i.e., $\neg \X \top$) is in
      $\negs(\P)$ and $\P$ is consistent, then it must be the case
      that $\X \top \in \pos(\P)$. We must have have $\E \END \in
      \sigma^+_2(\P)$, which means $\E \END \in
      \pos(\sigma(\P))$. It must be therefore be the case that
      $\E \END \in \pos(\Q)$ for all $\Q \in \comps(\sigma(\P))$,
      since each such $\Q$ must be an extension of $\sigma(\P)$.
  \end{proof}
\end{lemma}

\smallskip
We can go further, showing that $\E \END$ is in fact in \emph{every}
node's positive set, and every node is consistent and complete.

\begin{lemma}[Proof graphs are consistent]
  \label{lem:nodesconsistent} For all consistent and complete PNPs \P, every node $\Q \in \G_\P$
  is consistent and complete. If $\E \END \in \pos(\P)$, then $\E \END \in \pos(\Q)$.
  \begin{proof}
    By induction on the length of the shortest path from $\P$ to $\Q$
    in $\G_\P$\iffull\else, using Lemma~\ref{lem:endinvariant}\fi.
    When $n=0$, we have $\Q = \P$, so we have $\P$'s completeness
      and consistency by assumption; the second implication is immediate.

    When $n=n'+1$, we have some path
      $\P, \P_2, \P_3, \dots, \P_{n'}, \Q$. We know that $\P_{n'}$ is
      complete and consistent; we must show that $\Q$ is complete and
      consistent. By construction, we know that
      $\Q \in \comps(\sigma(\P_{n'}))$, so $\Q$ must be consistent and
      complete by definition. By the IH, we know that
      $\E \END \in \pos(\P_{n'})$, so by Lemma~\ref{lem:endinvariant}, we
      find the same for $\Q$.
  \end{proof}
\end{lemma}

Each node has the potential for successors: for each node $\Q \in
\G_P$, we can prove that $\Qhat$ implies that the disjunction of every
other node's literal interpretation holds in the next moment of time.

\begin{lemma}[Step implication]
  \label{lem:succprovable}
  For all consistent and complete PNPs $\P$ where $\E \END \in \pos(\P)$ then $\vdash \bigvee_{\Q \in \G_\P}
  \Qhat \Implies \WX \bigvee_{\Q \in \G_\P} \Qhat$.
  \begin{proof}
    Let $\Q \in \G_\P$ be given. We show that each $\Q$ implies the right-hand side.
    By Lemma~\ref{lem:transprovable}, we know that $\vdash \Qhat
    \Implies \WX \widehat{\sigma(\Q)}$.
    By Lemma~\ref{lem:nodesconsistent}, we know that $\Q$ is
    consistent and complete and $\E \END \in \pos(\Q)$. Since $\Q$ is
    complete, we know that $\END \in \F_\Q$.
    We now show that $\vdash \Qhat \Implies \WX \bigvee_{\Q' \in
      \comps(\sigma(\Q))} \Qhat'$, by cases on where $\END$ occurs in $\F_\Q$.

    If $\END \in \pos(Q)$), then $\comps(\sigma(\Q)) = \emptyset$ by
       Lemma~\ref{lem:endinvariant}; we must find
       $\vdash \Qhat \Implies \WX \bot$. Since
       $\vdash \Qhat \Implies \END$, we are done by
       Lemma~\ref{lem:nextwknext} with $\phi = \bot$.
    If, on the other hand, $\END \in \negs(Q)$, we have $\vdash \Qhat \Implies
      \neg \END$, so by Lemma~\ref{lem:transconsistent} we know that
      $\sigma(Q)$ is consistent. We therefore have $\vdash \widehat{\sigma(\Q)}
    \Implies \bigvee_{Q' \in \comps(\sigma(\Q))} \Qhat'$ by
    Lemma~\ref{lem:completionsprovable}.

    We have
    $\vdash \Qhat \Implies \WX \bigvee_{\Q' \in \G_\P} \Qhat'$,
    because $\comps(\sigma(\Q)) \subseteq \G_\P$ by definition. Since
    we find this for each $\Q$, we conclude
    $\vdash \bigvee_{\Q \in \G_\P} \Qhat \Implies \WX \bigvee_{\Q \in \G_\P} \Qhat$.
    \end{proof}
\end{lemma}

We have so far established that the proof graph $\G_\P$ is rooted at $\P$,
preserves any finiteness we may inject, and each node has provable successors. We are nearly
done: we show that our proof graph corresponds to a Kripke structure which models
$\P$.

\begin{definition}[Terminal nodes and paths]
  \label{def:terminalpaths}
  A node $\Z \in \G_\P$ is \emph{terminal} when $\X \top \in
  \negs(\Z)$. A path $\P_1, \dots, \P_n$ is
  \emph{terminal} when $\P_n$ is terminal.
\end{definition}

\begin{lemma}[Proof graphs are models]
  \label{lem:graphoperators}
  For all consistent and complete PNPs \P, if $\P_1, \P_1, \P_2,
  \dots, \P_n$ is a terminal path in $\G_\P$, then for
  all $i$:
  \begin{enumerate}
  \item For all formulae $\phi$, if $\X \phi \in \F_{\P_i}$ then $\X \phi
    \in \pos(\P_i)$ iff $\phi \in \pos(\P_{i+1})$.
  \item For all formulae $\phi$ and $\psi$, if $\phi \W \psi \in
    \F_{\P_i}$ then $\phi \W \psi \in \pos(\P_i)$ iff either $\phi \in
    \pos(P_j)$ for all $j \ge i$ or there is some $k \ge i$ such that
    $\psi \in \pos(\P_k)$ and $\forall i \le j < k, ~ \phi \in
    \pos(\P_j)$.
  \end{enumerate}
  \begin{proof} \mbox{}
    \begin{enumerate}
    \item We have $\P_{i+1} \in \comps(\sigma(\P_i))$ by
      definition. One the one hand, if $\X \phi \in \pos(\P_i)$, we have $\phi \in
        \pos(\sigma(\P_i))$, and so all consistent completions have
        $\phi$ in the positive set---in particular, $\P_{i+1}$.

      On the other hand, if we have $\phi \in \pos(\P_{i+1})$, 
        it must be the case that $\X \phi$ is in one of $\pos(\P_i)$
        or $\negs(\P_i)$ because $\X \phi \in \F_{P_i}$. In the former case, we are done
        immediately. Suppose for a contradiction that $\X \phi \in
        \negs(\P_i)$. Since $\P_{i+1}$ is a completion of
        $\sigma(\P_i)$, it must be that $\negs(\sigma(\P_i)) \subseteq
        \negs(\P_{i+1})$. Since $\X \phi \in \negs(\P_i)$, we must
        have $\phi \in \negs(\sigma(\P_i))$, so $\phi \in
        \negs(\P_{i+1})$. But we have $\phi \in \pos(\P_{i+1})$ by
        assumption---and we have contradicted the consistency of
        $\P_{i+1}$ (Lemma~\ref{lem:nodesconsistent}).
    \item We have $\P_{j+1} \in \comps(\sigma(\P_j))$ for all $j$, by
      definition. Further, we know that $\phi \W \psi \in \pos(\P_i)$
      implies that $\{ \phi, \psi \} \subseteq \F_{\P_i}$. We go by
      cases on where $\phi \W \psi$ occurs in $\F_{\P_i}$,
      going from left-to-right both times and proving the contrapositive for the right-to-left implication.
      \begin{enumerate}[align=left]
      \item[($\phi \W \psi \in \pos(\P_i)$)] We must show that either
        $\phi \in \pos(P_j)$ for all $j \ge i$ or there is some $k \ge
        i$ such that $\psi \in \pos(\P_k)$ and $\forall i \le j < k, ~
        \phi \in \pos(\P_j)$.
        We show (for all $\P$ on the path) that if $\phi \W \psi \in
        \pos(\P)$, then either $\psi \in \pos(\P)$ or $\phi \in
        \pos(\P)$ and for all $\Q \in \comps(\sigma(\P))$, we have
        $\phi \W \psi \in \pos(\Q)$.
        Since $\phi \W \psi \in \pos(\P)$, by \rn{WkUntilUnroll} we
        know that $\vdash \Phat \Implies \psi \vee \phi \wedge \WX
        (\phi \W \psi)$. Since $\psi$ and $\phi$ are both in $\F_\P$,
        we can simply inspect $\P$. If $\psi \in \pos(\P)$, we are
        done. So suppose $\psi \in \negs(\P)$. We must therefore have
        $\phi \in \pos(\P)$. By the definition of $\sigma^+_2$, we
        have $\phi \W \psi \in \sigma(\P)$, and so any $\Q \in
        \comps(\sigma(\P))$ must also have $\phi \W \psi \in \pos(\Q)$. \\

        We strengthen the inductive hypothesis, showing that for the
        remainder of the terminal path $\P_i \dots \P_{i+n}$
        either $\{ \phi, \phi \W \psi \} \subseteq \pos(\P_{j})$ for
        all $i \le j \le n$, or there exists a $k \ge i$ such that
        $\psi \in \pos(\P_{k})$ and $\{ \phi, \phi \W \psi \} \in
        \pos(\P_j)$ for all $i \le j < k$.
        We go by induction on $n$. When $n=0$, we either have $\psi \in
          \pos(\P_i)$ (and so $k = i$) or $\phi \in \pos(\P_i)$ (and
          then the path ends).
        When $n=n'+1$, we know the path from $\P_i$ to
          $\P_{i+n}$ has either $\phi$ in every positive set or
          eventually $\psi$ occurs after $\phi$s. In the latter case,
          we can simply reuse the $k$ from the inductive hypothesis.
          In the former case, we know  $\{ \phi, \phi \W \psi \}
          \subseteq \pos(\P_{n'})$, so by the above we can find that
          either $\psi \in \P_{n'}$ or since $\P_n \in
          \comps(\sigma(\P_{n'}))$ has $\phi \W \psi \in
          \pos(\P_n)$. By the above again, we can find that either
          $\psi \in \pos(\P_n)$ (and so $k=n$) or $\phi \in
          \pos(\P_n)$ (and we have $\phi \in \pos(\P_j)$ for all $j
          \ge i$).

      \item[($\phi \W \psi \not\in \pos(\P_i)$)] We have $\phi \W \psi
        \in \negs(\P_i)$, so we must show that it is not the case that
        either $\phi \in \pos(P_j)$ for all $j \ge i$ or there is some
        $k \ge i$ such that $\psi \in \pos(\P_k)$ and $\forall i \le j
        < k, ~ \phi \in \pos(\P_j)$.
        We show that all paths out of $\P_i$ have $\phi$ in the
        positive set for zero or more transitions, but eventually
        neither $\phi$ nor $\psi$ holds.

        First, we show that if $\phi \W \psi \in \negs(\P)$, then (a)
        $\psi \in \negs(\P)$ and (b) either $\phi \in \negs(\P)$ or $\phi \in
        \pos(\P)$ and $\forall \Q \in \comps(\sigma(\P))$, we have
        $\phi \W \psi \in \negs(\Q)$.
        Since $\phi \W \psi \in \negs(\P)$, we have $\vdash \Phat
        \Implies \neg (\psi \vee \phi \wedge \WX (\phi \W \psi))$ by
        \rn{WkUntilUnroll}. By \rn{Taut} we have $\vdash \Phat
        \Implies \neg \psi \wedge (\neg phi \vee \neg \WX (\phi \W
        \psi)$; by desugaring and $\rn{Taut}$ we have $\vdash \Phat
        \Implies \neg \psi \wedge (\neg \phi \vee \X \neg (\phi \W
        \psi)$.
        To have $\P$ consistent, it must be that $\psi \in
        \negs(\P)$. If $\phi \in \negs(\P)$, we are done---we have
        satisfied (a) and (b). Suppose $\phi \in \pos(\P)$. By the
        definition $\sigma^-_4$, we now have $\phi \W \psi \in
        \negs(\sigma(\P))$, so it must be the case that for any
        completion $\Q$, we have $\phi \W \psi \in \negs(\Q)$.

        Now, finally, suppose $\phi \W \psi \in \negs(\P_i)$. For
        $\P_i$ to be consistent, it must be the case that $\vdash
        \Phat_i \Implies \Phat_{i+1} \Implies \dots \Implies \Phat_n$
        (since no node is terminal until $\P_n$). One such node must
        have $\phi \in \negs(\P_i)$: apply the reasoning above to see
        that no node can have $\psi \in \pos(\P_i)$ and, furthermore,
        if $\phi \in \pos(\P_i)$ then $\phi \W \psi \in
        \negs(\P_{i+1})$. If it does not happen before the terminal
        node, the last one has no successor, so \rn{WkUntilUnroll}
        shows that necessarily $\phi \in \negs(\P_n)$. \qedhere
      \end{enumerate}
    \end{enumerate}
  \end{proof}
\end{lemma}
\smallskip
Here we slightly depart from Kr\"oger and Merz's presentation: since their
models can be infinite, they must make sure that their paths are
able to in some sense `fulfill' temporal predicates. We, on the other hand, know
that all of our paths are finite, so our reasoning is simpler. First,
there must exist some terminal node.
\begin{lemma}[Injected finiteness guarantees terminal nodes] ~\\
  \label{lem:terminalnodes}
  For all consistent and complete PNPs $\P$, if $\E \END \in \pos(\P)$
  then there is a terminal node $\Z \in \G_\P$.
  \begin{proof}
    Suppose for a contradiction that $\X \top \in
    \pos(Q)$ for all $\Q \in \G_\P$.

    We have assumed $\E \END \in \pos(\P)$; by desugaring, $\E \END$ amounts
    to $\neg \A \neg \END$, i.e., $\neg (\neg \END \W \bot)$, i.e.,
    $\neg (\X \top \W \bot)$.

    We have $\vdash \Q \Implies \X \top$ for each $\Q \in \G_\P$ by
    assumption, so by \rn{Taut}, we have $\vdash \bigvee_{\Q \in
      \G_\P} \Qhat \Implies \X \top$.

    We have $\vdash \bigvee_{\Q \in \G_\P} \Qhat \Implies \WX
    \bigvee_{\Q \in \G_\P}$ by Lemma~\ref{lem:succprovable}. So by
    \rn{Induction}, $\vdash \bigvee_{\Q \in \G_\P} \Implies \A \X
    \top$, i.e, $\X \top \W \bot$. Since $\P \in \G_\P$, we know
    $\vdash \Phat \Implies \X \top \W \bot$ by
    Lemma~\ref{lem:succprovable} again. But $\E \END \in \pos(\P)$
    means that $\vdash \Phat \Implies \E \END$, so $\vdash \Phat
    \Implies \neg (\X \top \W \bot)$, as well! It must then be the
    case that $\vdash \neg \Phat$, which contradicts our assumption
    that $\P$ is consistent.

    We therefore conclude that there must exist some node $\Z \in
    \G_\P$ such that $\X \top \in \negs(\Z)$.
  \end{proof}
\end{lemma}

Since our proof graph is constructed connectedly from the root on out,
the existence of a terminal node implies the existence of a terminal
path from the root to that node.

\smallskip

\begin{corollary}[Injected finiteness guarantees terminal paths] ~\\
  \label{cor:terminalpaths}
  For all consistent and complete PNPs $\P$, if $\E \END \in \pos(\P)$
  then there is a terminal path $\P, \P_2, \dots, \P_{n-1}, \Z \in \G_\P$.
  \begin{proof}
    By Lemma~\ref{lem:terminalnodes}, there exists some terminal node
    $\Z \in \G_\P$. Since $\G_\P$ is constructed by iterating $\comps$
    and $\sigma$ on $\P$, there must exist some $\P_{n-1}$ such that
    $\Z \in \comps(\sigma(\P_{n-1}))$, and some $\P_{n-2}$ such that
    $\P_{n-1} \in \comps(\sigma(\P_{n-2}))$ and so on back to
    $\P$---yielding a path.
  \end{proof}
\end{corollary}
We can now prove the key lemma: consistent PNPs are satisfiable in
their literal interpretation. A proof graph for a consistent PNP $\P$
induces a Kripke structure modeling $\P$'s literal interpretation,
$\Phat$. The proof actually considers a version of $\P$ with $\E \END$
(the \rn{Finite} axiom) injected into the positive set---we inject
finiteness to make sure we're building an appropriately finite model.

\begin{theorem}[\LTLf satisfiability]
  \label{thm:ltlfsatisfiability}
  If $\P$ is a consistent PNP, then $\Phat$ is satisfiable.
  \begin{proof}
    Let $\P' = (\{ \E \END \} \cup \pos(\P), \negs(\P))$. If $\P$ is
    consistent, then so is $\P'$. (If not, it must be because $\vdash
    \Phat \Implies \neg \E \END$; by \rn{Taut} and \rn{Finite}, we
    have $\vdash \Phat \Implies \E \END$, and so $\vdash \neg \Phat$
    and $\P$ is not consistent.)

    To show that $\Phat$ is satisfiable, we use the terminal path
    from Corollary~\ref{cor:terminalpaths} to construct a Kripke
    structure. Suppose our terminal path is of the form $\P, \P_2,
    \dots, \P_n$; let $\K^n = (\eta_1,\dots,\eta_n)$ where we define:
    \[ \eta_i(v) = \begin{cases}
      \true & v \in \pos(\P_i) \\
      \false & \text{otherwise}
    \end{cases}
    \]
    We must show that $\K^n_1(\Phat) = \true$; it suffices to show
    that $\K^n_1(\Phat') = \true$, since the variables in $\P$ and $\P'$ are identical.
    We prove that for all $\phi \in \F_{P'}$, we have $\K^n_i(\phi) = \true$ iff $\phi
    \in \pos(\P_i)$. We go by induction on $\phi$; throughout, we rely
    on the fact that every node is consistent and complete
    (Lemma~\ref{lem:nodesconsistent}).
    \begin{enumerate}[align=left]
    \item[($\phi=v$)] $\K^n_i(v) = \true$ iff $\eta_i(v) = \true$ iff $v \in \pos(\P_i)$.
    \item[($\phi=\bot$)] $\K^n_i(\bot) = \false$ by definition
      and $\bot \not\in \pos(\P_i)$ by Lemma~\ref{lem:pnpproperties}.
    \item[($\phi=\psi \Implies \chi$)] Let an $i$ be given. We know
      $\P_i$ is a consistent and complete PNP, so $\{ \psi, \chi \}
      \in F_{\P_i}$. By the IH, we have $\K^n_i(\psi) = \true$ iff
      $\psi \in \pos(\P_i)$ and similarly for $\chi$.
      We have $\K^n_i(\psi \Implies \chi) = \true$ iff $\K^n_i(\psi) =
      \false$ or $\K^n_i(\chi) = \true$ iff $\psi \in \negs(\P_i)$ or
      $\chi \in \pos(\P_i)$ (by the IHs) iff $\psi \Implies \chi \in \pos(\P_i)$\iffull{}
      (again by Lemma~\ref{lem:pnpproperties})\fi.
    \item[($\phi=\X \psi$)] Let an $i$ be given.  We have $\K^n_i(\X
      \psi) = \true$ iff $i > n$ and $\K^n_{i+1}(\psi) \true$ iff in
      $\K^n_{i+1}(\psi) = \true$ iff $\psi \in \pos(\P_{i+1})$ (by the
      IH) iff $\X \psi \in \pos(\P_i)$ (since $\X \psi \in \F_{\P'}$,
      by Lemma~\ref{lem:graphoperators}).
    \item[($\phi=\psi \W \chi$)] We have $\K^n_i(\psi \W \chi) = \true$
      iff either for all $i \le j \le n, \K^n_j(\psi) = \true$ or there
      exists a $i \le k \le n$ such that $\K^n_k(\chi) = \true$ and for
      all $i \le j < k$ we have $\K^n_j(\psi) = \true$. By the IH,
      those hold iff formulae are in appropriate positive sets; by
      Lemma~\ref{lem:graphoperators}, those formulae are in
      appropriate positive sets iff $\psi \W \chi$ is in the
      appropriate positive set.
    \end{enumerate}

    At this point, $\K^n_1(\Phat) = \true$ is a special case where $i=1$.
  \end{proof}
\end{theorem}
Finally, we can show completeness. The proof is the usual one, where
we to find a proof of $\vdash \phi$ we use $\models \phi$ to see that
$\neg \phi$ is unsatisfiable---and then the PNP for $\neg \phi$ is
inconsistent, and so $\vdash \neg \neg \phi$, which yields
$\vdash \phi$.

\begin{theorem}[\LTLf completeness]
  \label{thm:ltlfcompleteness}
  If $\models \phi$ then $\vdash \phi$.
  \begin{proof}
    If $\models \phi$, then for all Kripke structures $\K^n$, we have
    $\K^n_i(\phi) = \true$ for all $i$. Conversely, it must also be the
    case that $\K^n_i(\neg \phi) = \false$ for all $i$, and so $\neg
    \phi$ is unsatisfiable. In other words, the PNP $(\emptyset, \{
    \phi \})$ is unsatisfiable. By the contrapositive of
    Theorem~\ref{thm:ltlfsatisfiability}, it must be the case that
    $(\emptyset, \{ \phi \})$ is inconsistent, i.e., $\vdash \neg \neg
    \phi$. By \rn{Taut}, we can conclude that $\vdash \phi$.
  \end{proof}
\end{theorem}
We extend the proof of completeness to allow for assumptions in the
usual way.
\begin{corollary}[\LTLf completeness, with contexts]
  \label{cor:ltlfcompletenessctx}
  If $\F \models \phi$ then $\F \vdash \phi$.
  \begin{proof}
    By induction on the size of $\F$.
    If $|F| = 0$, then by Theorem~\ref{thm:ltlfcompleteness}.
    If $|F| = n + 1$, we have
      $\{ \phi_1, \dots, \phi_{n+1} \} \models \psi$. By
      Theorem~\ref{thm:semanticdeduction}, we have
      $\{ \phi_1, \dots, \phi_n \} \models \A \phi_{n+1} \Implies \psi$. By
      the IH, we have
      $\{ \phi_1, \dots, \phi_n \} \vdash \A \phi_{n+1} \Implies \psi$. By
      Theorem~\ref{thm:deduction}, we have
      $\{ \phi_1, \dots, \phi_{n+1} \} \vdash \psi$.
  \end{proof}
\end{corollary}

\section{Decision procedure}
\label{sec:decision}

We have implemented a satisfiability decision procedure
for \LTLf.\footnote{https://github.com/ericthewry/ltlf-decide} Our
method is based Kr\"oger and Merz's tableau-based decision
procedure~\cite{KrogerMerz_LTL_2008}. Kro\"ger and Merz generate
tableaux where the states are PNPs; they proceed to unfold
propositional and then temporal formulae while checking
for \emph{closedness}. If a certain kind of path exists in the
resulting graph, then the formula is satisfiable---we can use that
path to generate a Kripke structure.

The closed nodes of their tableaux are inductively defined as those
which are manifestly contradictory (e.g., $\bot \in \pos(\P)$ or
$\pos(\P) \cap \negs(\P) \ne \emptyset$), those where all of their
successors are contradictory (e.g. $\bot \W \bot \in \pos(P)$ isn't
obviously contradictory, but both of its temporal successors are), and
those where a negated temporal formula is never actually falsified
(e.g., if $\A \phi \in \negs(\P)$ and we are generating an infinite
Kripke structure, we had better falsify $\phi$ at some point).
The third criterion is a critical one: Kr\"oger and Merz, by default,
generate infinite paths in their tableaux, which correspond to
infinite Kripke structures. If they were to drop their third
criterion, they would find infinite paths where, say, $\neg \A \phi$
is meant to hold but $\phi$ is never falsified. Such ``dishonest''
infinite paths must be carefully avoided.

Our decision procedure diverges slightly from theirs. First, we
generalize their approach from just having always ($\A$) to include
weak until ($\W$). Next, we simplify their approach to exclude the
third condition on paths. Since we deal with finite models of time,
we'll never consider infinite paths---and so we avoid the issue of
dishonest infinite paths wholesale.

Our simplified notion of closedness means we can implement a more
efficient algorithm. While Kr\"oger and Merz need to keep the tableau
around in order to identify the ``honest'' strongly connected
components of the tableau, we need not do so. We can perform a
perfectly ordinary graph search without having to keep the whole
tableau in memory. (We do have to keep the \emph{states} of the
tableau in memory, though.) To be clear: we claim no asymptotic
advantage, and our algorithm remains exponential; rather, our
implementation is simpler.
We don't report on the efficiency of implementation at all---rather,
the code is written in Literate Haskell and is meant to be expository
and tutorial.

\section{Discussion}
\label{sec:discussion}

We have studied a finite temporal logic for linear time: \LTLf.
We were able to adapt techniques for \emph{infinite} temporal logics to show
deductive completeness in a finite setting.
We are by no means the first to prove completeness for \LTLf, but we
do so (a) in direct analogy to existing methods and (b) improving on
Ro\c{s}u's axioms~\cite{CoinductiveLTLf_2016}.
The proof of deductive completeness calls for only minor changes to
the proof with potentially infinite time: we \emph{inject finiteness}
by inserting $\E \END$ into our proof graphs, allowing us to directly
adapt methods from an infinite logic;
injecting finiteness simplifies the selection of the path used to
generate the Kripke structure in the satisfiability proof
(Lemma~\ref{lem:terminalnodes} and Corollary~\ref{cor:terminalpaths}).
We believe that the technique is general, and will adapt to other temporal
logics; we offer this proof as evidence.

To be clear, we claim that the proof of completeness for a `finitized'
logic is relatively straightforward \emph{once you find the right
  axioms}.
We can offer only limited guidance on finding the right axioms.
Finite temporal logics should have an axiom saying that time is, indeed, finite;
some sort of axiom will be needed to establish the meaning of temporal
modalities at the end of time (e.g., \rn{Finite});
when porting axioms from the infinite logic, one must be careful to
check that the axioms are sound at the end of time (e.g.,
\rn{EndNextContra}), when temporal modalities may change in meaning
(e.g., changing distribution over implication to use the weak next
modality, as in \rn{WkNextDistr}).

\section*{Acknowledgments}

The comments of anonymous FoSSaCS reviewers helped improve this work.

\bibliographystyle{asl}
\bibliography{ltl}

@inproceedings{CoinductiveLTLf_2016,
  title={Finite-Trace Linear Temporal Logic: Coinductive Completeness},
  author={Ro{\c{s}}u, Grigore},
  booktitle={{International Conference on Runtime Verification}},
  series = {RV '16},
  pages={333--350},
  year={2016},
  organization={Springer}
}

@inproceedings{TemporalNetKat_2016,
 author = {Beckett, Ryan and Greenberg, Michael and Walker, David},
 title = {Temporal NetKAT},
 booktitle = {{Programming Language Design and Implementation}},
 series = {PLDI '16},
 year = {2016},
 isbn = {978-1-4503-4261-2},
 location = {Santa Barbara, CA, USA},
 pages = {386--401},
 numpages = {16},
 url = {http://doi.acm.org/10.1145/2908080.2908108},
 doi = {10.1145/2908080.2908108},
 acmid = {2908108},
 publisher = {ACM},
 address = {New York, NY, USA},
}

@inproceedings{de2015synthesis,
 author = {De Giacomo, Giuseppe and Vardi, Moshe Y.},
 title = {Synthesis for LTL and LDL on Finite Traces},
 booktitle = {{International Joint Conference on Artificial Intelligence}},
 series = {IJCAI'15},
 year = {2015},
 isbn = {978-1-57735-738-4},
 location = {Buenos Aires, Argentina},
 pages = {1558--1564},
 numpages = {7},
 url = {http://dl.acm.org/citation.cfm?id=2832415.2832466},
 acmid = {2832466},
 publisher = {AAAI Press},
}

@article{PDL,
  title={Propositional dynamic logic of regular programs},
  author={Fischer, Michael J and Ladner, Richard E},
  journal={{Journal of Computer and System Sciences}},
  volume={18},
  number={2},
  pages={194--211},
  year={1979},
  publisher={Elsevier}
}

@inproceedings{DeGiacomo:2016:LFL:3060621.3060766,
 author = {Giuseppe De Giacomo and Moshe Y. Vardi},
 title = {{LTLf} and {LDLf} Synthesis Under Partial Observability},
 booktitle = {{International Joint Conference on Artificial Intelligence}},
 series = {IJCAI'16},
 year = {2016},
 isbn = {978-1-57735-770-4},
 location = {New York, New York, USA},
 pages = {1044--1050},
 numpages = {7},
 url = {http://dl.acm.org/citation.cfm?id=3060621.3060766},
 acmid = {3060766},
 publisher = {AAAI Press},
}

@inproceedings{Insensitivity_2014,
 author = {De Giacomo, Giuseppe and De Masellis, Riccardo and Montali, Marco},
 title = {Reasoning on LTL on Finite Traces: Insensitivity to Infiniteness},
 booktitle = {{National Conference on Artificial Intelligence}},
 series = {AAAI'14},
 year = {2014},
 location = {Qu\&\#233;bec City, Qu\&\#233;bec, Canada},
 pages = {1027--1033},
 numpages = {7},
 url = {http://dl.acm.org/citation.cfm?id=2893873.2894033},
 acmid = {2894033},
 publisher = {AAAI Press},
}

@inproceedings{NetKat_2014,
 author = {Anderson, Carolyn Jane and Foster, Nate and Guha, Arjun and Jeannin, Jean-Baptiste and Kozen, Dexter and Schlesinger, Cole and Walker, David},
 title = {NetKAT: Semantic Foundations for Networks},
 booktitle = {{Symposium on Principles of Programming Languages}},
 series = {POPL '14},
 year = {2014},
 isbn = {978-1-4503-2544-8},
 location = {San Diego, California, USA},
 pages = {113--126},
 numpages = {14},
 url = {http://doi.acm.org/10.1145/2535838.2535862},
 doi = {10.1145/2535838.2535862},
 acmid = {2535862},
 publisher = {ACM},
 address = {New York, NY, USA},
}

@inproceedings{IntroLDLf_2012,
  title={Linear temporal logic and linear dynamic logic on finite traces},
  author={De Giacomo, Giuseppe and Vardi, Moshe Y},
  booktitle={{International Joint Conference on Artificial Intelligence}},
  pages={854--860},
  year={2013},
  organization={Association for Computing Machinery}
}

@book{KrogerMerz_LTL_2008,
       place={Berlin},
       title={Temporal logic and state systems},
       publisher={Springer},
       author={Kröger, Fred and Merz, Stephan},
       year={2008},
       annotate={ This book presents an overview of Temporal logic and its
                  extensions, construcing LTL and its well-studied extensions
                  almost from first Logical principles. It is a text intended
                  for those well-versed in Logics and Meta-Mathematics, and will
                  inform much of the semantic and proof-theoretic analysis of
                  Linear Temporal Logic over finite Traces. The proof of
                  Soundness and Completeness for LTL over finite traces will
                  likely be very similar to the proof given in this work. }}

@inproceedings{PastTimeLTL_1985,
  title={The Glory of the Past},
  author={Lichtenstein, Orna and Pnueli, Amir and Zuck, Lenore},
  booktitle={{Workshop on Logic of Programs}},
  pages={196--218},
  year={1985},
  organization={Springer}
}

@INPROCEEDINGS{Concurrent_1977, 
               author={A. Pnueli}, 
               booktitle={{Foundations of Computer Science}}, 
               title={The temporal logic of programs}, 
               year={1977}, 
               pages={46-57}, 
               doi={10.1109/SFCS.1977.32},
	       ISSN={0272-5428},
	       month={Oct},
	       annotate={This provides a formulation of Program Correctness
                  proof using Temporal Logics that forms the basis of the
                  Curry-Howard Correspondence between LTL and FRP}}

@inproceedings{Baier:2006:PFT:1597538.1597664,
 author = {Baier, Jorge A. and McIlraith, Sheila A.},
 title = {Planning with First-order Temporally Extended Goals Using Heuristic Search},
 booktitle = {{National Conference on Artificial Intelligence}},
 series = {AAAI'06},
 year = {2006},
 isbn = {978-1-57735-281-5},
 location = {Boston, Massachusetts},
 pages = {788--795},
 numpages = {8},
 url = {http://dl.acm.org/citation.cfm?id=1597538.1597664},
 acmid = {1597664},
 publisher = {AAAI Press},
}

@inproceedings{DAntoni:2017:MSL:3009837.3009844,
 author = {D'Antoni, Loris and Veanes, Margus},
 title = {Monadic Second-order Logic on Finite Sequences},
 booktitle = {{Symposium on Principles of Programming Languages}},
 series = {POPL 2017},
 year = {2017},
 isbn = {978-1-4503-4660-3},
 location = {Paris, France},
 pages = {232--245},
 numpages = {14},
 url = {http://doi.acm.org/10.1145/3009837.3009844},
 doi = {10.1145/3009837.3009844},
 acmid = {3009844},
 publisher = {ACM},
 address = {New York, NY, USA},
}

@article{Kozen97kat,
  author    = {Dexter Kozen},
  title     = {Kleene Algebra with Tests},
  journal   = {{ACM} Trans. Program. Lang. Syst.},
  volume    = {19},
  number    = {3},
  pages     = {427--443},
  year      = {1997},
  url       = {http://doi.acm.org/10.1145/256167.256195},
  doi       = {10.1145/256167.256195},
  bibsource = {dblp computer science bibliography, http://dblp.org}
}

\end{document}